\documentclass[english,12pt]{article}
\usepackage[T1]{fontenc}
\usepackage[latin9]{inputenc}
\usepackage{geometry}
\usepackage{color}
\usepackage{babel}
\usepackage{float}
\usepackage{booktabs}
\usepackage{mathtools}
\usepackage{amsmath}
\usepackage{amsthm}
\usepackage{amssymb}
\usepackage{graphicx}
\usepackage{setspace}
\usepackage[authoryear]{natbib}
\usepackage[unicode=true,pdfusetitle,
 bookmarks=true,bookmarksnumbered=false,bookmarksopen=false,
 breaklinks=false,pdfborder={0 0 0},pdfborderstyle={},backref=false,colorlinks=true]
 {hyperref}
\hypersetup{
 citecolor=blue}

\makeatletter

\providecommand{\tabularnewline}{\\}
\floatstyle{ruled}
\newfloat{algorithm}{tbp}{loa}
\providecommand{\algorithmname}{Algorithm}
\floatname{algorithm}{\protect\algorithmname}

\theoremstyle{plain}
\newtheorem{assumption}{\protect\assumptionname}
\theoremstyle{plain}
\newtheorem{thm}{\protect\theoremname}
\theoremstyle{plain}
\newtheorem{cor}{\protect\corollaryname}

\usepackage{algorithmic}

\usepackage{setspace}

\usepackage[compact]{titlesec}

\date{}

\@ifundefined{showcaptionsetup}{}{%
 \PassOptionsToPackage{caption=false}{subfig}}
\usepackage{subfig}
\makeatother

\providecommand{\assumptionname}{Assumption}
\providecommand{\corollaryname}{Corollary}
\providecommand{\theoremname}{Theorem}

\begin{document}
\addtolength{\abovedisplayskip}{-5.75pt}
\addtolength{\belowdisplayskip}{-5.75pt}
\renewcommand{\abstractname}{}

\title{\textbf{Gradient-based Sparse Principal Component Analysis with Extensions
to Online Learning}}
\author{Yixuan Qiu, Jing Lei, and Kathryn Roeder\\
{\normalsize{}Department of Statistics and Data Science, Carnegie
Mellon University}\\
{\normalsize{}Pittsburgh, PA 15213, }\texttt{\normalsize{}\{yixuanq,
jinglei, roeder\}@andrew.cmu.edu}}
\maketitle
\begin{abstract}
\noindent \textbf{Abstract}: Sparse principal component analysis
(PCA) is an important technique for dimensionality reduction of high-dimensional
data. However, most existing sparse PCA algorithms are based on non-convex
optimization, which provide little guarantee on the global convergence.
Sparse PCA algorithms based on a convex formulation, for example the
Fantope projection and selection (FPS), overcome this difficulty,
but are computationally expensive. In this work we study sparse PCA
based on the convex FPS formulation, and propose a new algorithm that
is computationally efficient and applicable to large and high-dimensional
data sets. Nonasymptotic and explicit bounds are derived for both
the optimization error and the statistical accuracy, which can be
used for testing and inference problems. We also extend our algorithm
to online learning problems, where data are obtained in a streaming
fashion. The proposed algorithm is applied to high-dimensional gene
expression data for the detection of functional gene groups.\medskip{}

\end{abstract}

\section{Introduction}

Principal component analysis (PCA, \citealp{pearson1901liii,hotelling1933analysis})
is a classical yet indispensable dimensionality reduction technique
in statistics and machine learning. PCA generates higher-level features
of the raw data by computing uncorrelated linear combinations of the
original variables that retain the maximum amount of variation of
the raw data. Moreover, PCA can process data sets that have a variable
dimension larger than the sample size. Such desirable properties of
PCA make it one of the most popular preprocessing techniques in multivariate
statistics.

In the high-dimensional setting where the number of variables can
be comparable to or larger than the sample size, PCA suffers from
the well-known curse-of-dimensionality. For instance, \citet{johnstone2009consistency}
and \citet{jung2009pca} showed that when the number of variables
is much larger than the sample size, PCA can behave poorly in estimating
the principal components (PCs), even with a simple population covariance
structure, producing misleading results in scenarios that it was exactly
invented for.

On the other hand, these theoretical works also motivated the development
of a variant of PCA, the sparse PCA method, which overcame many of
the limitations of traditional PCA in high-dimensional settings. Sparse
PCA works similarly to the original PCA, but requires the PCs to be
sparse. Here sparsity means that the linear combination involves only
a small number of variables. Such a sparsity requirement greatly reduces
the number of coefficients to estimate, and enhances the interpretability
of the estimated PCs. Pioneer works on sparse PCA include \citet{jolliffe2003modified,johnstone2009consistency,zou2006sparse}
etc.. Since then sparse PCA has found wide applications in keyword
extraction for text data \citep{zhang2011large}, fault detection
for industrial processes \citep{grbovic2012decentralized,gajjar2018real},
genomics and genetics \citep{lee2012sparse,zhu2017testing}, among
many others.

One major challenge of sparse PCA is the computation. Unlike ordinary
PCA, which can be efficiently solved using well-studied eigen decomposition
methods such as the power method, the original formulation of sparse
PCA \citep{jolliffe2003modified} involves solving a sparsity constrained
eigenvalue problem that is computationally hard. Existing fast algorithms
for nonconvex objective functions \citep{zou2006sparse,witten2009penalized,journee2010generalized}
generally do not guarantee the global convergence and rely on the
initial values. This limitation has an adverse impact on the applications
of sparse PCA, especially in rigorous statistical inference and scientific
research. Alternatively, \citet{d2005direct,vu2013fantope} proposed
convex formulations of the sparse PCA problem using semidefinite programming,
which are computationally expensive for large matrices commonly seen
in modern applications such as text mining and bioinformatics. Therefore,
a sparse PCA algorithm that has both a global convergence guarantee
and an efficient implementation is in great need.

The computational difficulties of the existing sparse PCA algorithms
also limit their applications in an important area: the online learning
methods that arise from the demand to analyze large-scale streaming
data. As the volumes of data sets are rapidly growing and data collection
procedures become more dynamic, it is challenging to store and analyze
all the observations at the same time, so it is preferable to build
and update models immediately after a new data point is obtained.
Online PCA algorithms have been extensively studied in the literature
\citep{oja1985stochastic,warmuth2008randomized,marinov2018streaming,li2018near},
but the work on online sparse PCA is scarce \citep{yang2015streaming,wang2016online}.
The difficulty of online sparse PCA mainly comes from the fact that
existing methods could not express sparse PCA as an easy-to-solve
optimization problem. A statistically and computationally provable
online sparse PCA algorithm remains an open problem.

To overcome the challenges above, in this article we propose new computational
algorithms for sparse PCA and its online versions. The main contributions
of our work are as follows. First, by analyzing the geometry of sparse
PCA, we represent its solution by an \emph{unconstrained} convex optimization
problem. As a result, efficient gradient-based and projection-free
algorithms are developed, whose output can be used as good initial
values for nonconvex methods. Second, the unconstrained convex formulation
is extended to the online setting, and two different online sparse
PCA algorithms are proposed, depending on whether the data sets have
large sample sizes or high dimensions. To our best knowledge, these
are the first online sparse PCA algorithms that can be computed efficiently
and have global convergence guarantees for a general covariance model.
Third, for each algorithm, both the optimization error and the statistical
accuracy are rigorously analyzed with nonasymptotic and explicit bounds.

The theoretical justifications are supported by various simulation
experiments. For the batch version of sparse PCA, we demonstrate that
our new algorithm has much faster convergence than the existing method
given the same computational time. In online settings, the proposed
methods also have convergence results that are consistent with the
theory. Moreover, we apply the new sparse PCA algorithm to a real
high-dimensional gene expression data set and successfully detect
differential co-expression patterns in schizophrenia subjects compared
to a control group. Proofs of theorems are given in the supplementary
material.

\section{Overview of Sparse PCA}

\label{sec:overview}

From a statistical point of view, the major target of PCA is to estimate
the factor loadings of each PC from the noisy data. Suppose the data
set is a sample of independent and identically distributed random
vectors $Z_{1},\ldots,Z_{n}\in\mathbb{R}^{p}$ with zero means and
the true covariance matrix $\Sigma=Cov(Z_{i})$. Let $\theta_{i}=\theta_{i}(A),i=1,\ldots p$
represent the ordered eigenvalues of a matrix $A$, $\theta_{1}\ge\cdots\ge\theta_{p}$,
and $\gamma_{i}(A)$ be the associated eigenvector. Then PCA aims
at estimating the $p\times d$ matrix $\Gamma=(\gamma_{1}(\Sigma),\ldots,\gamma_{d}(\Sigma))$
containing the top $d$ eigenvectors of $\Sigma$, which is typically
referred to as the factor loading matrix.

The ordinary PCA estimates $\Gamma$ by first computing the sample
covariance matrix, $S=n^{-1}\sum_{i=1}^{n}Z_{i}Z_{i}^{\mathrm{T}}$,
and then extracting the leading $d$ eigenvectors of $S$. However,
it has been well studied that in the high-dimensional case $p\gg n$,
$S$ can be a poor estimator for $\Sigma$, so the ordinary PCA method
is also likely to fail. To enable PCA in high-dimensional data, one
needs to make stronger assumptions on the data distribution. For example,
in sparse PCA, $\Gamma$ is assumed to contain many zero entries,
so that the number of unknown coefficients are greatly reduced. This
idea leads to the following core assumption throughout this article.
\begin{assumption}
\label{assu:sparsity}The factor loading matrix $\Gamma$ has at most
$s$ nonzero rows, and the $d$-th eigengap of $\Sigma$ is nonzero,
$\delta_{d}=\theta_{d}(\Sigma)-\theta_{d+1}(\Sigma)>0$.
\end{assumption}
Such a sparsity assumption has been considered as the ``row sparsity''
in \citet{vu2013minimax}, which assumes that the leading $d$-dimensional
principal subspace is unique and is supported on a small number of
coordinates. This is a quite strong assumption, but in many applications
such as genetics, a sparse factor loading vector is often preferred
due to the better interpretability. Assumption \ref{assu:sparsity}
is made to facilitate the mathematical investigation of sparse PCA
algorithms.

Under the sparsity assumption, sparse PCA has been formulated in many
different ways, including the lasso approach in PCA \citep{jolliffe2003modified},
regression-based formulation \citep{zou2006sparse}, iterative thresholding
methods \citep{shen2008sparse,witten2009penalized,ma2013sparse,she2017selective},
the generalized power method \citep{journee2010generalized}, among
many others. Also see \citet{zou2018selective} for a recent review
of various sparse PCA methods. Despite the rich literature, most of
the existing algorithms suffer from two common issues. The first issue
is from the perspective of optimization. The majority of the existing
sparse PCA algorithms are formulated as nonconvex optimization problems,
which possess some local convergence properties at best. Therefore,
such algorithms highly rely on the initial values, which are typically
unavailable \emph{a priori}. The second issue is on the statistical
aspect. To recover the true population eigenvectors, sparse PCA methods
typically impose some additional structural assumptions on the covariance
matrix, for instance the spiked covariance model.

In comparison, convex optimization has the advantage of superior convergence
properties. In most cases, a proper algorithm can iteratively find
the global optimum irrespective of the initial values. Such a property
makes convex optimization extremely popular in statistical and machine
learning models. For sparse PCA, \citet{d2005direct} proposed a formulation
called DSPCA that takes the form of a convex semidefinite program.
Let $\Vert A\Vert_{p,q}=\{\sum_{j=1}^{n}(\sum_{i=1}^{m}|a_{ij}|^{p})^{q/p}\}^{1/q}$
denote the $L_{p,q}$ norm for an $m\times n$ matrix $A$, and then
DSPCA finds an estimator for the projection matrix $\Pi_{1}=\gamma_{1}\gamma_{1}^{\mathrm{T}}$
using the solution to the following optimization problem:
\begin{align}
\max\quad & \mathrm{tr}(SX)\nonumber \\
\mathrm{s.t.}\quad & \mathrm{tr}(X)=1,\ \Vert X\Vert_{1,1}\le s_{1},\ \text{and}\ O\preceq X,\label{eq:dspca_formulation}
\end{align}
where $s_{1}$ is a parameter to control the sparsity of the solution,
$O$ is the zero matrix, and $A\preceq B$ means $B-A$ is nonnegative
definite.

Since DSPCA only extracts the first component, \citet{vu2013fantope}
developed a generalized model, called Fantope projection and selection
(FPS), to estimate the top-$d$ projection matrix $\Pi=\Gamma\Gamma^{\mathrm{T}}$.
The optimization problem of FPS is given by
\begin{align}
\max\quad & \mathrm{tr}(SX)-\lambda\Vert X\Vert_{1,1}\nonumber \\
\mathrm{s.t.}\quad & O\preceq X\preceq I\text{ and }\mathrm{tr}(X)=d,\label{eq:fps_formulation}
\end{align}
where $\lambda$ is the sparsity penalty parameter. The convex constraint
set $\mathcal{F}^{d}=\{X:O\preceq X\preceq I\text{ and }\mathrm{tr}(X)=d\}$
is called the Fantope. When $d=1$, FPS becomes equivalent to DSPCA.
The FPS formulation has attractive statistical properties \citep{vu2013fantope,lei2015sparsistency},
and can be solved in polynomial time using the alternating direction
method of multipliers (ADMM, \citealp{boyd2011distributed}), an iterative
algorithm for constrained convex optimization problems.

However, the existing ADMM-based FPS algorithm is shown to be slow,
since each iteration of the algorithm requires projecting a $p\times p$
matrix onto the Fantope, which involves a full eigen decomposition
of the $p\times p$ matrix. When the dimensionality of $S$ is high,
for example in genetic studies, the computational cost of the ADMM
algorithm is $\mathcal{O}(p^{3})$ per iteration. As a consequence,
the applicability of FPS is substantially limited by the cubic growth
of computing time per iteration, and a more computationally efficient
FPS algorithm is much desired.

\section{A New Projection-Free Algorithm for Sparse PCA}

\subsection{Gradient-based Methods for Large-scale Optimization}

In convex optimization problems, if the objective function is twice
differentiable, then the standard approach is the Newton--Raphson
iteration based on the Hessian matrix. However, when the parameter
dimension is too high so that the Hessian matrix is too large, or
when the objective function is not differentiable, one often needs
to resort to the first-order methods that rely only on the gradient
or subgradient of the objective function. In this article we refer
to such methods as the gradient-based methods.

The gradient-based methods have successful applications in many statistical
and machine learning problems, but their computational efficiency
heavily depends on the form of the optimization problem. Take the
FPS problem (\ref{eq:fps_formulation}) as an example, which has two
difficulties to deal with. First, the objective function is nonsmooth,
and second, the solution is sought within a constrained set $\mathcal{F}^{d}$.
If one ignores the nonsmoothness, then a simple gradient-based method
is the projected subgradient descent algorithm,
\begin{equation}
X_{k+1}=\mathcal{P}_{\mathcal{F}^{d}}\left(X_{k}+\alpha_{k}S-\alpha_{k}\lambda\cdot\mathrm{sign}(X)\right),\label{eq:projected_gradient_fps}
\end{equation}
where $\alpha_{k}$ is the step size at iteration $k$, and the symbol
$\mathcal{P}_{C}(x)=\arg\min_{y\in C}\,\Vert y-x\Vert$ means the
projection of $x$ onto a convex set $C$, with $\Vert\cdot\Vert$
being the Euclidean norm. In (\ref{eq:projected_gradient_fps}), the
sign function $\mathrm{sign}(X)$ is the subgradient of the nonsmooth
$\Vert X\Vert_{1,1}$ term. To overcome the nonsmoothness, a faster
optimization scheme is given by the ADMM algorithm using proximal
operators, where the proximal operator of a convex function $f$ with
step size $\alpha$ is defined as $\mathbf{prox}_{\alpha f}(x)=\arg\min_{u}\left\{ f(u)+(2\alpha)^{-1}\Vert u-x\Vert^{2}\right\} $,
and can be seen as a special gradient. Let $\mathcal{S}_{\alpha}(x)=\mathrm{sign}(x)\cdot\max\{\vert x\vert-\alpha,0\}$
be the soft-thresholding operator, and $\mathcal{S}_{\alpha}(X)=\mathbf{prox}_{\alpha\Vert\cdot\Vert_{1,1}}(X)$
means applying $\mathcal{S}_{\alpha}(x)$ to the matrix $X$ elementwisely.
Then the ADMM algorithm proceeds as follows \citep{vu2013fantope},
\begin{align}
X_{k+1} & =\mathcal{P}_{\mathcal{F}^{d}}(Y_{k}-U_{k}+\alpha S),\label{eq:admm_fps}\\
Y_{k+1} & =\mathcal{S}_{\alpha\lambda}(X_{k+1}+U_{k}),\quad U_{k+1}=U_{k}+X_{k+1}-Y_{k+1},\nonumber 
\end{align}
where $Y$ and $U$ are auxiliary variables, and $\alpha$ is the
step size.

For both (\ref{eq:projected_gradient_fps}) and (\ref{eq:admm_fps}),
however, the projection operator $\mathcal{P}_{\mathcal{F}^{d}}$
is unavoidable, which becomes the major bottleneck of the overall
algorithms. Therefore, to accelerate the convex sparse PCA, it is
necessary to reformulate the objective function and get rid of the
time-consuming projection operator.

\subsection{Projection-free Optimization on Intersection of Convex Sets}

\label{subsec:opt_complex_sets}

The massive cost of $\mathcal{P}_{\mathcal{F}^{d}}$ stems from the
complexity of the constraint set $\mathcal{F}^{d}$, which is the
intersection of three convex sets: $\mathcal{F}_{1}=\{X:\mathrm{tr}(X)=d\}$,
$\mathcal{F}_{2}=\{X:X\succeq O\}$, and $\mathcal{F}_{3}=\{X:X\preceq I\}$.
Each one of the three sets has a simple structure. However, when taking
the intersection, the associated projection operator becomes the major
obstacle for an efficient algorithm.

To this end, in this section we first develop a general scheme for
solving optimization problems on the intersection of convex sets.
We show that under certain assumptions, the complex constraint can
be recast as a penalty term added to the objective function, so that
the original constrained optimization problem is equivalent to an
unconstrained one. Moreover, under a proper setting, the new problem
can bypass the complicated operators on the intersection set, and
directly work on each individual convex set, which significantly reduces
the computational difficulty.

The optimization problem considered in this section has the following
form:
\begin{equation}
\min_{x\in\mathcal{K}}\,f(x),\quad\mathcal{K}=C_{1}\cap\cdots\cap C_{l}\cap G_{1}\cap\cdots\cap G_{m},\label{eq:opt_formulation}
\end{equation}
where $f(x)$ is a convex function, $C_{i}$'s are closed convex sets,
and $G_{i}$ is defined by $G_{i}=\{x:g_{i}(x)\le0\}$. Each constraint
function $g_{i}(x)$ is a convex function, and $\mathcal{K}$ is contained
in a closed convex set $\mathcal{X}\subset\mathbb{R}^{p}$ whose projection
operator $\mathcal{P}_{\mathcal{X}}$ is trivial. The intersection
set $\mathcal{K}$ is decomposed in such a way that the projection
operators $\mathcal{P}_{C_{i}}$ and the constraint functions $g_{i}(x)$
are easy to compute.

The problem with $m=0$ has been studied in the literature \citep{kundu2018convex},
but it is not useful for the FPS problem since $\mathcal{P}_{\mathcal{F}_{2}}$
and $\mathcal{P}_{\mathcal{F}_{3}}$ are still expensive. As will
be shown in the next section, the inclusion of the $G_{i}$ sets overcomes
this difficulty, since the constraint functions only involve the extreme
eigenvalues of $X$. The problem with $l=0$ and $m=1$ has been studied
in \citet{mahdavi2012stochastic} and \citet{yang2017richer}. Obviously,
our formulation in (\ref{eq:opt_formulation}) is a generalization
to the ones mentioned above. We then make the following assumptions
on the objects involved in (\ref{eq:opt_formulation}).
\begin{assumption}
\label{assu:f_lipschitz}$f(x)$ is Lipschitz continuous on $\mathcal{X}$
with the Lipschitz constant $L>0$: $|f(x)-f(y)|\le L\Vert x-y\Vert$,
$\forall x,y\in\mathcal{X}$.
\end{assumption}
\begin{assumption}
\label{assu:g_cond}For $i=1,\ldots,m$, (a) $x\in\mathcal{X}$ implies
$\mathcal{P}_{G_{i}}(x)\in\mathcal{X}$; (b) there exists a constant
$\rho_{i}$ such that
\[
\inf_{\substack{x\in\bar{G}_{i}\cap\mathcal{X}\\
v\in\partial g_{i}(x)
}
}\Vert v\Vert\ge\rho_{i}>0,
\]
where $\bar{G}_{i}=\{x:g_{i}(x)=0\}$, and $\partial g_{i}(x)=\{v:g_{i}(y)-g_{i}(x)\ge v^{\mathrm{T}}(y-x),\forall y\}$
is the subdifferential of $g_{i}$ at $x$.
\end{assumption}
\begin{assumption}
\label{assu:gamma_h}There exist a constant $\gamma>0$ and a function
$h:[0,+\infty)^{l+m}\mapsto[0,+\infty)$ such that (a) $h(\mathbf{0})=0$,
(b) $h$ is nondecreasing in each argument, and (c) for all $x\in\mathcal{X}$,
\begin{equation}
d_{\mathcal{K}}(x)\le\gamma h\left(d_{C_{1}}(x),\ldots,d_{C_{l}}(x),d_{G_{1}}(x),\ldots,d_{G_{m}}(x)\right),\label{eq:bounded_intersection_distance}
\end{equation}
where $\mathbf{0}$ is the zero vector, and $d_{C}(x)=\Vert x-\mathcal{P}_{C}(x)\Vert$
is the distance between $x$ and $C$.
\end{assumption}
Assumption \ref{assu:f_lipschitz} is a common condition for objective
functions. Assumption \ref{assu:g_cond} is derived from \citet{yang2017richer},
and can also be easily verified given concrete $g_{i}(x)$ functions.
Assumption \ref{assu:gamma_h} is the key to transforming problem
(\ref{eq:opt_formulation}) into an unconstrained one, and to a great
extent it needs to be analyzed case by case. Verifying Assumption
\ref{assu:gamma_h} for the FPS problem is the main focus of Section
\ref{subsec:gradient_fps}. Define the function
\[
\mathcal{L}(x;\mu)=f(x)+\mu h\left(d_{C_{1}}(x),\ldots,d_{C_{l}}(x),\rho_{1}^{-1}[g_{1}(x)]_{+},\ldots,\rho_{m}^{-1}[g_{m}(x)]_{+}\right),
\]
where $[x]_{+}=\max\{x,0\}$. Then the following theorem, which can
be seen as a generalization to Proposition 2 of \citet{kundu2018convex},
states the equivalence between (\ref{eq:opt_formulation}) and an
unconstrained optimization problem $\min_{x\in\mathcal{X}}\,\mathcal{L}(x;\mu)$.
\begin{thm}
\label{thm:unconstrained_problem}Suppose that Assumptions \ref{assu:f_lipschitz}
to \ref{assu:gamma_h} hold, and define $f_{*}=\min_{x\in\mathcal{K}}\,f(x)$
and $\mathcal{L}_{*}=\min_{x\in\mathcal{X}}\,\mathcal{L}(x;\mu)$.
Also let $x_{\varepsilon}\in\mathcal{X}$ be an approximate solution
such that $\mathcal{L}(x_{\varepsilon};\mu)\le\mathcal{L}_{*}+\varepsilon$
for $\varepsilon>0$, and denote $y_{\varepsilon}=\mathcal{P}_{\mathcal{K}}(x_{\varepsilon})$.
Then the following conclusions hold: (a) if $\mu\ge\gamma L$, then
$f_{*}=\mathcal{L}_{*}$; (b) if $\mu\ge\gamma(L+1)$, then $\Vert x_{\varepsilon}-y_{\varepsilon}\Vert\le\varepsilon$
and $\mathcal{L}(y_{\varepsilon};\mu)\le\mathcal{L}_{*}+\varepsilon$.
\end{thm}

\subsection{The Gradient FPS Algorithm}

\label{subsec:gradient_fps}

The FPS problem (\ref{eq:fps_formulation}) can be written in the
form of (\ref{eq:opt_formulation}) by defining $f(X)=-\mathrm{tr}(SX)+\lambda\Vert X\Vert_{1,1}$,
$C_{1}=\{X:\mathrm{tr}(X)=d\}$, $g_{1}(X)=\theta_{1}(X)-1$, $g_{2}(X)=-\theta_{p}(X)$,
$G_{1}=\{X:g_{1}(X)\le0\}$, $G_{2}=\{X:g_{2}(X)\le0\}$, $\mathcal{K}=\mathcal{F}^{d}$,
and $\mathcal{X}=\{X_{p\times p}:\Vert X\Vert_{F}\le\sqrt{d}\}$,
where $\Vert\cdot\Vert_{F}\equiv\Vert\cdot\Vert_{2,2}$ is the Frobenius
norm. In the remaining part of this article, the above symbols are
specific to the FPS model. To apply Theorem \ref{thm:unconstrained_problem},
we need to verify the three assumptions presented in Section \ref{subsec:opt_complex_sets},
among which Assumption \ref{assu:gamma_h} plays a central role in
developing the unconstrained optimization problem. The following theorem,
which describes the geometry of the Fantope, is the key to validating
that assumption.
\begin{thm}
\label{thm:dist_ineq}Let $\mathcal{F}_{1}=\{X_{p\times p}:\mathrm{tr}(X)=d\}$
and $\mathcal{F}_{2,3}=\{X_{p\times p}:O\preceq X\preceq I\}$. If
$3\le d\le(p-1)/2$, then for any $p\times p$ symmetric matrix $X$,
\begin{equation}
d_{\mathcal{F}^{d}}(X)\le\sqrt{p/(d+1)}\cdot d_{\mathcal{F}_{1}}(X)+\sqrt{p}\cdot d_{\mathcal{F}_{2,3}}(X).\label{eq:dist_ineq}
\end{equation}
\end{thm}
Theorem \ref{thm:dist_ineq} is proved using the theory of normal
cones in convex analysis. With inequality (\ref{eq:dist_ineq}), we
are able to verify the required assumptions in the following corollary.
\begin{cor}
\label{cor:fps_constants}For the FPS problem (\ref{eq:fps_formulation}),
if $3\le d\le(p-1)/2$, then
\begin{enumerate}
\item $f(X)$ satisfies Assumption \ref{assu:f_lipschitz} with $L=\Vert S\Vert_{F}+\lambda p$.
\item Assumption \ref{assu:g_cond} holds with $\rho_{1}=1/\sqrt{d}$ and
$\rho_{2}=1/\sqrt{p}$.
\item $d_{\mathcal{K}}(X)\le\sqrt{p/(d+1)}\left(d_{C_{1}}(X)+\sqrt{d+1}\cdot d_{G_{1}}(X)+\sqrt{d+1}\cdot d_{G_{2}}(X)\right)$.
\end{enumerate}
As a consequence, define
\begin{equation}
\mathcal{L}(X)=-\mathrm{tr}(SX)+\lambda\Vert X\Vert_{1,1}+\mu\left(d_{C_{1}}(X)+r_{1}[g_{1}(X)]_{+}+r_{2}[g_{2}(X)]_{+}\right),\label{eq:unconstrained_objective}
\end{equation}
and then $\min_{X\in\mathcal{K}}\,f(X)=\min_{X\in\mathcal{X}}\,\mathcal{L}(X)$,
where $\mu=(L+1)\sqrt{p/(d+1)}$, $r_{1}=\sqrt{d(d+1)}$, and $r_{2}=\sqrt{p(d+1)}$.
\end{cor}
Since projection onto $\mathcal{X}$ is trivial, (\ref{eq:unconstrained_objective})
is essentially an unconstrained objective function, which can be minimized
using any familiar subgradient method. However, subgradient methods
for nonsmooth objective functions in general require $\mathcal{O}(1/\varepsilon^{2})$
iterations to achieve an optimization error of $\varepsilon$, which
may be slow in practice. Below we introduce an efficient algorithm
that only requires $\mathcal{O}(1/\varepsilon)$ outer iterations.
For convenience, define $f_{1}(X)=\lambda\Vert X\Vert_{1}$ and $f_{2}(X)=-\mathrm{tr}(SX)+\mu d_{C_{1}}(X)+\mu r_{1}[g_{1}(X)]_{+}+\mu r_{2}[g_{2}(X)]_{+}$,
so the problem becomes $\min_{X\in\mathcal{X}}\,\mathcal{L}(X)\coloneqq f_{1}(X)+f_{2}(X)$.
Then we apply the proximal-proximal-gradient method \citep{ryu2017proximal},
which evaluates the proximal operators for $f_{1}$ and $f_{2}$ iteratively.
The outline of the proposed method, which we term as the gradient
FPS algorithm, or GradFPS for short, is given in Algorithm \ref{alg:grad_fps}.

\begin{algorithm}
\caption{\label{alg:grad_fps}The gradient FPS (GradFPS) algorithm}

\begin{spacing}{1.5}

\begin{algorithmic}[1]

\REQUIRE $S$, $T$, $\alpha$, initial value $X_{0}\in\mathcal{X}$

\ENSURE $\hat{X}$

\STATE $Z_{0}^{(1)}=Z_{0}^{(2)}\leftarrow X_{0}$

\FOR{ $k=0,1,\ldots,T-1$ }

\STATE $\bar{Z}_{k}\leftarrow(Z_{k}^{(1)}+Z_{k}^{(2)})/2$

\STATE $X_{k+1}\leftarrow\mathcal{P}_{\mathcal{X}}\left(\bar{Z}_{k}\right)=\min\left\{ 1,\sqrt{d}/\Vert\bar{Z}_{k}\Vert_{F}\right\} \cdot\bar{Z}_{k}$

\STATE $Z_{k+1}^{(1)}\leftarrow Z_{k}^{(1)}-X_{k+1}+\mathbf{prox}_{\alpha f_{1}}(2X_{k+1}-Z_{k}^{(1)})$

\STATE $Z_{k+1}^{(2)}\leftarrow Z_{k}^{(2)}-X_{k+1}+\mathbf{prox}_{\alpha f_{2}}(2X_{k+1}-Z_{k}^{(2)})$

\ENDFOR

\RETURN $\hat{X}=T^{-1}\sum_{k=1}^{T}X_{k}$

\end{algorithmic}

\end{spacing}
\end{algorithm}

We comment that the operations in Algorithm \ref{alg:grad_fps} are
all inexpensive compared with a full eigen decomposition. First, $\mathbf{prox}_{\alpha f_{1}}(X)=\mathcal{S}_{\alpha\lambda}(X)$
is the elementwise soft-thresholding operator, which has a closed-form
solution. We provide two algorithms for computing the proximal operator
for $f_{2}$: one is a direct method, and the other is an iterative
method. The details of the two algorithms are given in Appendix \ref{subsec:computation_prox_f2}.

\subsection{Convergence Analysis}

One remarkable benefit of the GradFPS algorithm is that we can bound
its optimization error at any finite iteration step. With a sufficiently
large number of iterations, Algorithm \ref{alg:grad_fps} can be shown
to output an $\varepsilon$-optimal and $\varepsilon$-feasible solution
$\hat{X}$, in the sense that $\mathcal{L}(\hat{X})\le\mathcal{L}_{*}+\varepsilon$
and $d_{\mathcal{K}}(\hat{X})\le\varepsilon$. We develop the convergence
property and an explicit upper bound for the optimization error in
the following theorem.
\begin{thm}
\label{thm:batch_fps_convergence}The output $\hat{X}$ of Algorithm
\ref{alg:grad_fps} satisfies
\[
\mathcal{L}(\hat{X})\le\min_{X\in\mathcal{X}}\,\mathcal{L}(X)+\frac{C}{T}\quad\text{and}\quad d_{\mathcal{K}}(\hat{X})\le\frac{C}{T},
\]
where $C$ is a constant that only depends on $S$, $X_{0}$, and
the model parameters. The explicit expression of $C$ is given in
Appendix \ref{subsec:constants}.
\end{thm}
If the optimization problem $\min_{X\in\mathcal{X}}\,\mathcal{L}(X)$
can be solved exactly, resulting in a solution $\hat{X}_{*}$, then
the statistical property of $\hat{X}_{*}$ has already been studied
by \citet{vu2013fantope}. However, in any practical implementation,
only a finite-precision solution such as $\hat{X}$ can be obtained.
$\hat{X}$ differs from the ideal $\hat{X}_{*}$ in two aspects: it
does not exactly minimize the objective function, and it is not necessarily
within the constraint set $\mathcal{K}$. In Corollary \ref{cor:stat_error},
we show that despite the presence of such approximations, $\hat{X}$
is still a good estimator for $\Pi$, and we explicitly give an upper
bound of its estimation error as a function of the sample size $n$
and the number of iterations $T$.
\begin{assumption}
\label{assu:samp_cov_sub_exponential}There exists a constant $\sigma>0$
such that $\max_{i,j}\,P(|S_{ij}-\Sigma_{ij}|\ge u)\le2\exp(-4nu^{2}/\sigma^{2})$
for all $u\le\sigma$.
\end{assumption}
\begin{cor}
\label{cor:stat_error}Suppose that Assumptions \ref{assu:sparsity}
and \ref{assu:samp_cov_sub_exponential} hold, and take $\lambda=\sigma\sqrt{\log(p)/n}$.
Then with probability at least $1-2/p^{2}$, we have
\begin{equation}
\Vert\hat{X}-\Pi\Vert_{F}\le\frac{4\sigma s\sqrt{\log(p)}}{\delta_{d}\sqrt{n}}+\frac{\sqrt{2C/\delta_{d}}}{\sqrt{T}}+\frac{C}{T},\label{eq:estimation_error}
\end{equation}
where $C$ is given in Theorem \ref{thm:batch_fps_convergence}.
\end{cor}
The error bound (\ref{eq:estimation_error}) has an intuitive interpretation.
The first term quantifies the \emph{statistical error}, which depends
on the $\log(p)$ term that is common in high-dimensional data analysis.
The second term is the \emph{optimization error}, which decays at
the $\mathcal{O}(1/\sqrt{T})$ rate. The last term is the \emph{feasibility
error}, since $\hat{X}$ is not necessarily a projection matrix.

\section{Online Sparse PCA}

\label{sec:online_sparse_pca}

\subsection{Online Learning Setting}

In this section we consider the scenario in which data are obtained
in a streaming fashion. Streaming data reflect many practical needs
that data acquisition and computation happen roughly at the same time.
For instance, the complete data collection procedure may span a long
period of time, or the data set is too large to be stored entirely
on the machine. In both cases, it is desirable to make full use of
the existing data, and then update the model parameters when new data
points come in. Such algorithms are typically called online learning
algorithms. Correspondingly, the algorithms that use the whole data
set, for instance Algorithm \ref{alg:grad_fps}, are referred to as
offline learning or batch learning algorithms.

Formally, we assume that there is an infinite sequence of independent
random vectors $Z_{1},Z_{2},\ldots\in\mathbb{R}^{p}$ with $E(Z_{t})=0$
and $Cov(Z_{t})=E(S_{t})=\Sigma$, $t\ge1$, where $S_{t}=Z_{t}Z_{t}^{\mathrm{T}}$.
The true covariance matrix $\Sigma$ has the same sparsity setting
as the batch version, and the estimation target is the top-$d$ projection
matrix $\Pi$ of $\Sigma$. We define the online learning procedure
as follows. At each time point $t$, the data analyst constructs an
estimator $X_{t}$ for $\Pi$. To match the nature of streaming data,
we require that $X_{t}$ only depends on $Z_{t}$, $X_{t-1}$, and
optionally some other quantities that depend on the history $\{Z_{i}\}_{i=0}^{t}$
with a storage size not growing with $t$. The procedure stops at
time $T$, and a final estimator $\hat{X}_{T}$ is output by the online
learning algorithm. For clarity, $T$ is also called the sample size
of the streaming data in this context.

The performance of an online algorithm is evaluated based on both
the statistical and optimization properties. For the final output
$\hat{X}_{T}$, we are interested in its estimation error $\Vert\hat{X}_{T}-\Pi\Vert_{F}$.
And for the whole estimator sequence $\{X_{t}\}$, we also care about
its cumulative optimization loss $\mathcal{R}(\{X_{t}\},T)$, defined
in the following way. After each $X_{t}$ is constructed, we use it
to predict a future data point $Z_{t+1}$, and define the loss function
\begin{equation}
\ell_{t}(X_{t})=-Z_{t+1}^{\mathrm{T}}X_{t}Z_{t+1}+\lambda\Vert X_{t}\Vert_{1,1}+\nu d_{\mathcal{K}}(X_{t}),\label{eq:loss_function_definition}
\end{equation}
where $\lambda$ and $\nu$ are constants. In this loss function,
the first term quantifies the (negative) explained variance on new
data if $X_{t}$ is treated as a projection matrix, the second term
encourages the sparsity of $X_{t}$, and the third term penalizes
the deviation from the constraint set $\mathcal{K}=\mathcal{F}^{d}$.
For the whole procedure, define the total loss
\begin{equation}
\mathcal{R}(\{X_{t}\},T)=\sum_{t=1}^{T}\ell_{t}(X_{t})-\sum_{t=1}^{T}\ell_{t}(\Pi),\label{eq:regret_definition}
\end{equation}
which describes the cumulative excess loss of $\{X_{t}\}$ compared
with the true projection matrix $\Pi$. In online learning literature,
the function $\mathcal{R}(\cdot)$ is typically called the \emph{regret}.
Naturally, a good online learning algorithm should have a strict control
of the regret as a function of $T$. In the next two sections, we
propose two different online sparse PCA algorithms based on the characteristics
of the streaming data.

\subsection{The Large-sample-size Case}

\label{subsec:high_frequency_case}

The first case is the typical setting of streaming data, where new
data are obtained with a high frequency. As a result, the sample size
$T$ is assumed to be much larger than the dimension $p$. The primary
goal of the online learning algorithm is to make quick prediction
$X_{t}$ after the data point $Z_{t}$ is observed, and meanwhile
to control the regret and final estimation error.

Under this setting, we solve the online sparse PCA problem using the
incremental proximal method \citep{bertsekas2011incremental}, which
is a generalization to the simple subgradient method. Originally designed
for batch optimization problems, the incremental proximal method is
extended to the online setting in this article. We call the proposed
algorithm Online-T GradFPS, to indicate that it is mostly used for
streaming data that have a large sample size $T$. The outline of
Online-T GradFPS is given in Algorithm \ref{alg:prox_online_fps}.

\begin{algorithm}
\caption{\label{alg:prox_online_fps}The Online-T GradFPS algorithm}

\begin{spacing}{1.5}

\begin{algorithmic}[1]

\REQUIRE $\{Z_{t}\}$, $T$, $\{\alpha_{t}\}$, initial value $X_{0}$

\ENSURE $\hat{X}_{T}$

\FOR{ $t=1,\ldots,T$ }

\STATE $X_{t}^{(0)}\leftarrow X_{t-1}$

\STATE $X_{t}^{(1)}\leftarrow\mathcal{S}_{\alpha_{t}\lambda}(X_{t}^{(0)})$

\STATE $X_{t}^{(2)}\leftarrow X_{t}^{(1)}-\alpha_{t}\nu\sqrt{pd}\mathbf{1}\{\theta_{1}>1\}\gamma_{1}\gamma_{1}^{\mathrm{T}}+\alpha_{t}\nu p\mathbf{1}\{\theta_{p}<0\}\gamma_{p}\gamma_{p}^{\mathrm{T}}$,\\
 where $\theta_{i}=\theta_{i}(X_{t}^{(1)})$, $\gamma_{i}=\gamma_{i}(X_{t}^{(1)})$,
$i=\{1,p\}$

\STATE $X_{t}^{(3)}\leftarrow X_{t}^{(2)}+\min\{\beta,1\}\cdot s\cdot I$,
where $s=(d-\mathrm{tr}(X_{t}^{(2)}))/p$, $\beta=\alpha_{t}\nu/\{(d+1)|s|\}$

\STATE $X_{t}\leftarrow\mathcal{P}_{\mathcal{X}}\left(X_{t}^{(3)}+\alpha_{t}S_{t}\right)=\min\left\{ 1,\sqrt{d}/\Vert X_{t}^{(3)}+\alpha_{t}S_{t}\Vert_{F}\right\} \cdot\left(X_{t}^{(3)}+\alpha_{t}S_{t}\right)$

\ENDFOR

\RETURN $\hat{X}_{T}=T^{-1}\sum_{t=1}^{T}X_{t}$

\end{algorithmic}

\end{spacing}
\end{algorithm}

Compared with Algorithm \ref{alg:grad_fps}, Online-T GradFPS has
a significantly lower computational cost per iteration, due to the
following two reasons. First, the eigenvalues are computed for a sparse
matrix $X_{t}^{(1)}$, since it is the output of a soft-thresholding
operator. Computing the extreme eigenvalues for $X_{t}^{(1)}$ is
much more efficient than for a dense matrix, since its complexity
depends on the number of nonzero elements. Second, only the largest
and smallest eigenvalues of $X_{t}^{(1)}$ need to be calculated,
which further saves the computation time.

The following theorem shows that if $\Vert S_{t}\Vert_{F}$ is properly
bounded, then the average regret of Algorithm \ref{alg:prox_online_fps}
decays at the rate of $\mathcal{O}(1/\sqrt{T})$, which matches the
best known result for the online subgradient method on a non-strongly
convex objective function.
\begin{assumption}
\label{assu:prox_online_fps_assumption}(a) The sequence $\xi_{t}=\Vert S_{t}-\Sigma\Vert_{F}$,
$t\ge1$ is independent and identically distributed, with a sub-exponential
distribution. (b) The sequence $\zeta_{t}=\Vert S_{t}\Vert_{F}^{2}$
is also sub-exponential. Specifically, there exist constants $b_{1},b_{2},\sigma_{1},\sigma_{2}\ge0$
such that
\begin{align*}
E\left[\exp\{\lambda(\xi_{t}-\mu_{1})\}\right]\le\exp(\lambda^{2}\sigma_{1}^{2}/2) & ,\quad\forall\,|\lambda|\le1/b_{1},\\
E\left[\exp\{\lambda(\zeta_{t}-\mu_{2})\}\right]\le\exp(\lambda^{2}\sigma_{2}^{2}/2) & ,\quad\forall\,|\lambda|\le1/b_{2},
\end{align*}
where $\mu_{1}=E(\xi_{t})$ and $\mu_{2}=E(\zeta_{t})$.
\end{assumption}
\begin{thm}
\label{thm:prox_online_fps_convergence}Let $\alpha_{1}=\alpha_{0}>0$
and $\alpha_{t}=\alpha_{0}/\sqrt{t-1}$ for $t\ge2$. Then the following
conclusions hold:
\begin{enumerate}
\item \textup{(Optimization regret bound)} If $\Vert S_{t}\Vert_{F}$ is
bounded, then $T^{-1}\mathcal{R}(\{X_{t}\},T)=\mathcal{O}(p^{2}/\sqrt{T})$.
\item \textup{(Statistical estimation error)} If Assumptions \ref{assu:sparsity}
and \ref{assu:prox_online_fps_assumption} hold, and $\nu\ge\lambda p+\Vert\Sigma\Vert_{F}+1$,
then for any fixed $\varepsilon\in(0,1)$,
\[
\Vert\hat{X}_{T}-\Pi\Vert_{F}=\mathcal{O}\left(\sqrt{(\log(1/\varepsilon)+\nu^{2}p^{2})/\sqrt{T}+\lambda s}\right)
\]
holds with probability at least $1-\varepsilon$.
\end{enumerate}
The explicit expressions of $\mathcal{R}(\{X_{t}\},T)$ and $\Vert\hat{X}_{T}-\Pi\Vert_{F}$
are given in Appendix \ref{subsec:constants}.
\end{thm}
Theorem \ref{thm:prox_online_fps_convergence} indicates that $\lambda$
needs to be set small if the primary goal is to use the final output
$\hat{X}_{T}$ for estimation. Otherwise, a moderate $\lambda$ leads
to more sparse intermediate results and is thus better for interpretation.
The estimation error bound also implies that Online-T GradFPS has
a slower convergence rate than the batch GradFPS with respect to the
sample size. However, as has been explained previously, the major
advantage of Online-T GradFPS is its computational efficiency, which
offsets its weakness in estimation error.

\subsection{The High-dimensional Case}

\label{subsec:high_dimension_case}

When the data dimension $p$ is much larger than the sample size $T$,
the method in Section \ref{subsec:high_frequency_case} is no longer
applicable, since both the regret of $\{X_{t}\}$ and the estimation
error of $\hat{X}_{T}$ depend on a polynomial of $p$. As a comparison,
in high-dimensional statistical analysis, such quantities usually
depend on $\log(p)$ under suitable sparsity assumptions. Therefore,
we are motivated to consider alternative optimization schemes that
result in a smaller regret and a better statistical accuracy, possibly
at the expense of larger computational cost in each iteration.

For general online learning problems, one of the most natural and
straightforward methods to obtain $X_{t}$ is to apply the batch algorithm
on all collected data $S_{1},\ldots,S_{t}$ up to time $t$. Such
a scheme is known as the follow-the-leader (FTL) algorithm. For online
sparse PCA, the FTL algorithm is a valid online learning algorithm,
since the matrix $S_{1:t}=t^{-1}\sum_{i=1}^{t}S_{i}$ can be computed
with a constant storage. However, the main problem of FTL is its weak
control of the regret, as the numerical experiment shows in Section
\ref{sec:simulation}. Intuitively, FTL focuses too much on the existing
data, and leaves little room for the exploration of future observations.

Instead, we develop our online sparse PCA algorithm based on the generalized
online mirror descent framework (OMD, \citealp{orabona2015generalized}).
The key merit of the generalized OMD method is to replace the Frobenius
norm $\Vert S_{t}\Vert_{F}$ in the error bound (\ref{eq:prox_online_fps_regret})
by the infinity norm $\Vert S_{t}\Vert_{\infty,\infty}$, which only
grows at the speed of $\log(p)$ under some regularity conditions.
Due to this reason, the proposed algorithm is named as Online-P GradFPS,
to emphasize that it is more suitable for a large $p$.

For brevity, we set the constants $r=\log(p)/\{\log(p)-1\}$ and $\beta=\exp(-4)/\{\log(p)-1\}$,
and define the function
\begin{align}
\mathring{\mathcal{L}}(X;Y,t)= & -\mathrm{tr}(YX)+\lambda t\Vert X\Vert_{1,1}+\sqrt{t}\Vert X\Vert_{r,r}^{2}/2\nonumber \\
 & +(L_{t}+1)\sqrt{p/(d+1)}\left(d_{C_{1}}(X)+r_{1}[g_{1}(X)]_{+}+r_{2}[g_{2}(X)]_{+}\right),\label{eq:l_mirror}
\end{align}
where $L_{t}=\Vert Y\Vert_{F}+\lambda tp+\exp(-4)\sqrt{td}p^{2}$.
We reuse the notation in Section \ref{subsec:gradient_fps} for other
terms in (\ref{eq:l_mirror}). The main steps of Online-P GradFPS
are given in Algorithm \ref{alg:mirror_online_fps}. It is worth mentioning
that we improve the original OMD method by allowing an approximate
solution for the subproblem in each iteration (line 4 of Algorithm
\ref{alg:mirror_online_fps}), which is more realistic and efficient
in practice. Solving the subproblem of Algorithm \ref{alg:mirror_online_fps}
is very similar to that of Algorithm \ref{alg:grad_fps}, and we provide
the details in Appendix \ref{subsec:subproblem_omd}.

\begin{algorithm}
\caption{\label{alg:mirror_online_fps}The Online-P GradFPS algorithm}

\begin{spacing}{1.5}

\begin{algorithmic}[1]

\REQUIRE $\{S_{t}\}$, $\{\varepsilon_{t}\}$, $T$, $\lambda$

\ENSURE $\hat{X}$

\STATE $Y_{0}\leftarrow O$

\FOR{ $t=1,\ldots,T$ }

\STATE $Y_{t}\leftarrow Y_{t-1}+S_{t}$

\STATE Find $X_{t}\in\mathcal{X}$ such that $\mathring{\mathcal{L}}(X_{t};Y_{t},t)\le\mathring{\mathcal{L}}_{*}+\beta\sqrt{t}\varepsilon_{t}^{2}/2$,\\
where $\mathring{\mathcal{L}}_{*}=\min_{X\in\mathcal{X}}\,\mathring{\mathcal{L}}(X;Y_{t},t)$

\ENDFOR

\RETURN $\hat{X}_{T}=X_{T}$

\end{algorithmic}

\end{spacing}
\end{algorithm}

Similar to the large-sample-size case, the following theorem describes
both the optimization error and the statistical accuracy of Algorithm
\ref{alg:mirror_online_fps}.
\begin{thm}
\label{thm:mirror_online_fps_convergence}The following conclusions
hold:
\begin{enumerate}
\item \textup{(Optimization regret bound) If $\Vert S_{t}\Vert_{\infty,\infty}$
is bounded and $\varepsilon_{t}=\mathcal{O}(1/\sqrt{t})$, then }$T^{-1}\mathcal{R}(\{X_{t}\},T)=\mathcal{O}(1/\sqrt{T})$.
\item \textup{(Statistical estimation error)} If Assumptions \ref{assu:sparsity}
and \ref{assu:samp_cov_sub_exponential} hold, $\lambda=\sigma\sqrt{\log(p)/T}$,
and $\varepsilon_{t}=1/\sqrt{t}$, then with probability at least
$1-2/p^{2}$, we have
\[
\Vert\hat{X}_{T}-\Pi\Vert_{F}=\mathcal{O}\left(\frac{s\sqrt{\log(p)}+s^{2-4/\log(p)}}{\sqrt{T}}\right).
\]
\end{enumerate}
The explicit expressions of $\mathcal{R}(\{X_{t}\},T)$ and $\Vert\hat{X}_{T}-\Pi\Vert_{F}$
are given in Appendix \ref{subsec:constants}.
\end{thm}
Comparing the results in Theorem \ref{thm:mirror_online_fps_convergence}
and those in Theorem \ref{thm:prox_online_fps_convergence}, it is
clear that the optimization errors of the two algorithms have the
same order for $T$, but the estimation error of Online-P GradFPS
decays faster than that of Online-T GradFPS. The price for the better
estimation accuracy is a larger computational cost per iteration,
which will be made clear in the simulation study.

\section{Simulation Study}

\label{sec:simulation}

\subsection{Simulation Setting}

\label{subsec:simulation_setting}

\begin{figure}[h]
\begin{centering}
\subfloat[]{\includegraphics[width=0.45\textwidth]{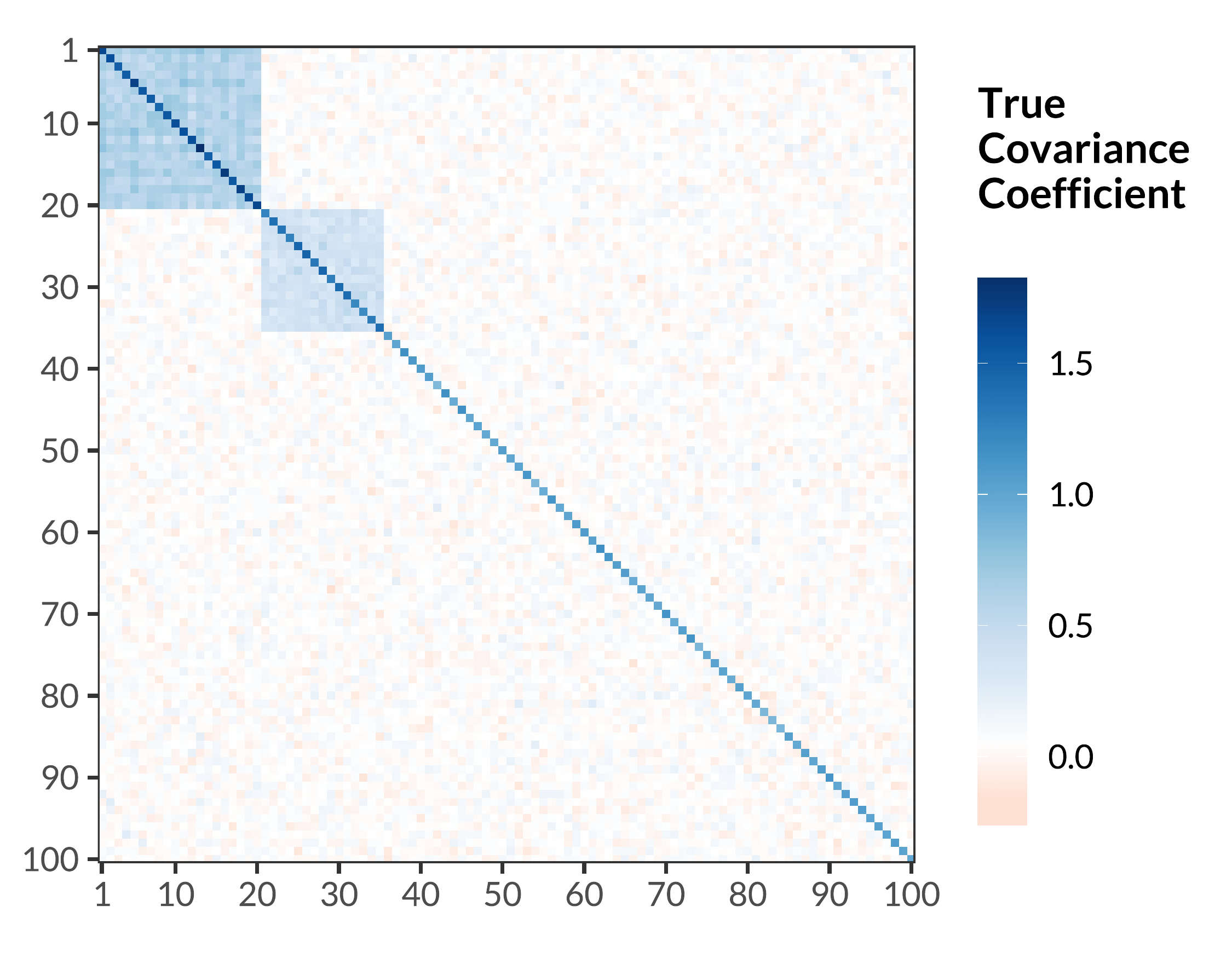}

}\subfloat[]{\includegraphics[width=0.55\textwidth]{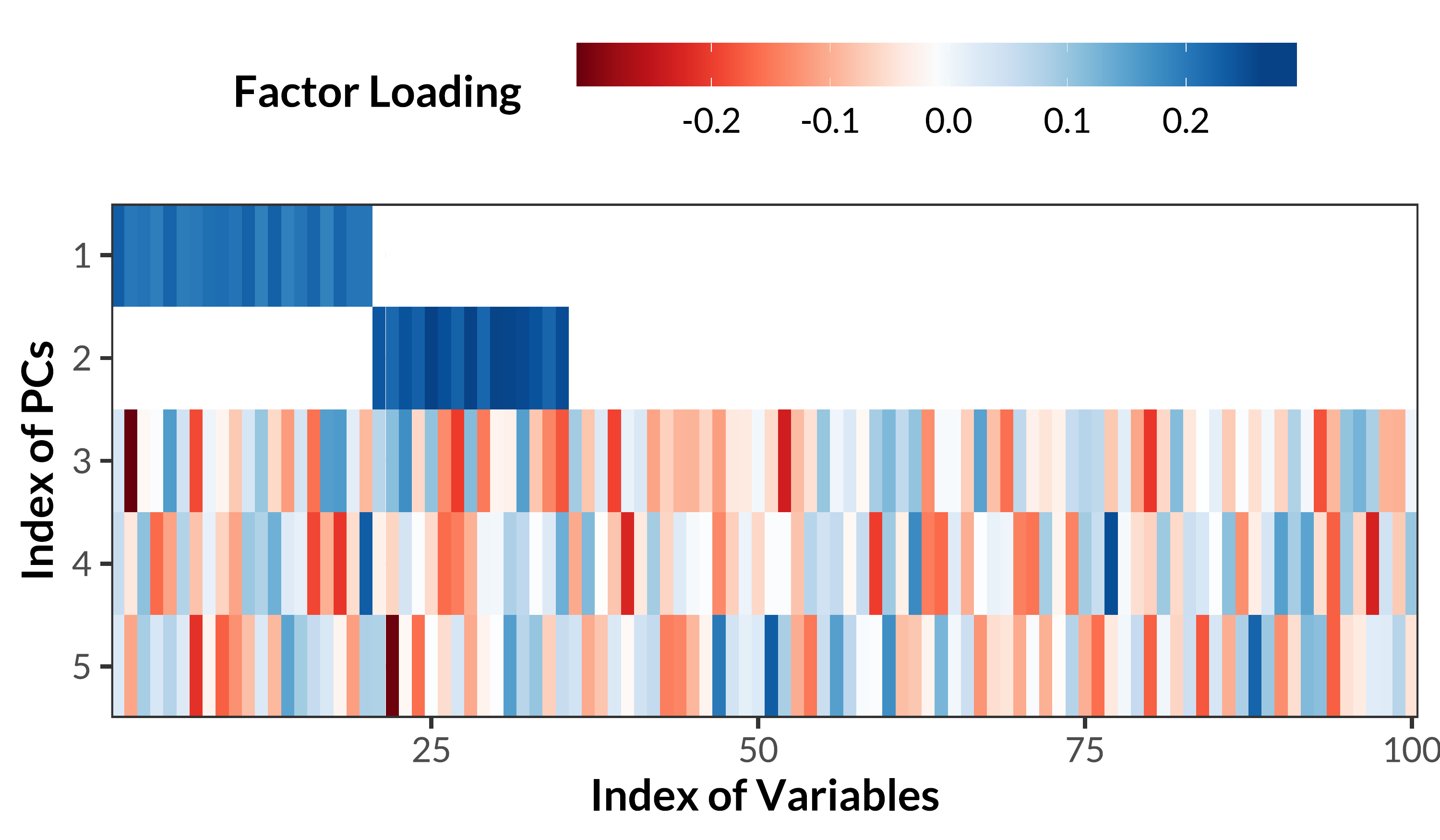}

}
\par\end{centering}
\caption{\label{fig:simulation_setting}(a) The true covariance matrix $\Sigma$
with $p=100$. (b) The eigenvectors of $\Sigma$ associated with the
five largest eigenvalues.}
\end{figure}

In this section we conduct a number of numerical experiments to evaluate
the performance of the sparse PCA algorithms proposed in this article.
The problem setting is as follows. We assume that the data $Z_{1},\ldots,Z_{n}$
follow independent and identically distributed multivariate normal
distribution $N(0,\Sigma)$, where $\Sigma$ is the true covariance
matrix of $p$ variables, and $n$ is the sample size. For online
learning algorithms, the data sequence is of infinite length, and
the online algorithm will choose a terminal sample size $T$. The
$p$ variables are categorized into three groups: the first signal
group contains $d_{1}=20$ variables, the second signal group contains
$d_{2}=15$ variables, and the last noise group consists of $(p-d_{1}-d_{2})$
noise variables. Figure \ref{fig:simulation_setting}(a) gives a visualization
of the true covariance matrix $\Sigma$ with $p=100$, which shows
that most variables are weakly correlated with each other, but the
ones within the same signal group have higher correlations. In different
experiments, $n$ and $p$ may vary, but $d_{1}$ and $d_{2}$ are
kept fixed.

The $\Sigma$ matrix is obtained by generating the eigenvalues $\Lambda$
and eigenvectors $Q$ in the following way. Let $U_{r_{1}:r_{2},c_{1}:c_{2}}$
denote the submatrix of a $p\times p$ matrix $U$, with row indices
$r_{1}$ to $r_{2}$ and column indices $c_{1}$ to $c_{2}$. When
$r_{1}=r_{2}$ or $c_{1}=c_{2}$, a single index is used. First simulate
a $U$ matrix such that $U_{1:d_{1},1}\overset{iid}{\sim}Unif(0.9,1.1)$,
$U_{(d_{1}+1):p,1}=0$, $U_{1:d_{1},2}=U_{(d_{1}+d_{2}+1):p,2}=0$,
$U_{(d_{1}+1):(d_{1}+d_{2}),2}\overset{iid}{\sim}Unif(0.9,1.1)$,
and $U_{1:p,3:p}\overset{iid}{\sim}N(0,1)$. Then a QR decomposition
is performed as $U=QR$, and $Q$ is used as the eigenvectors of $\Sigma$.
Next, let $\Lambda=\mathrm{diag}\{12,6,\lambda_{3},\ldots,\lambda_{p}\}$,
where $\lambda_{i}\overset{iid}{\sim}Unif(0,2)$, and then $\Sigma$
is computed as $\Sigma=Q\Lambda Q^{\mathrm{T}}$. Figure \ref{fig:simulation_setting}(b)
shows the first five columns of $Q$, and clearly the first $d=2$
columns of $Q$ contain the sparse eigenvectors.

\subsection{Batch Algorithms}

\begin{figure}
\begin{centering}
\includegraphics[width=0.95\textwidth]{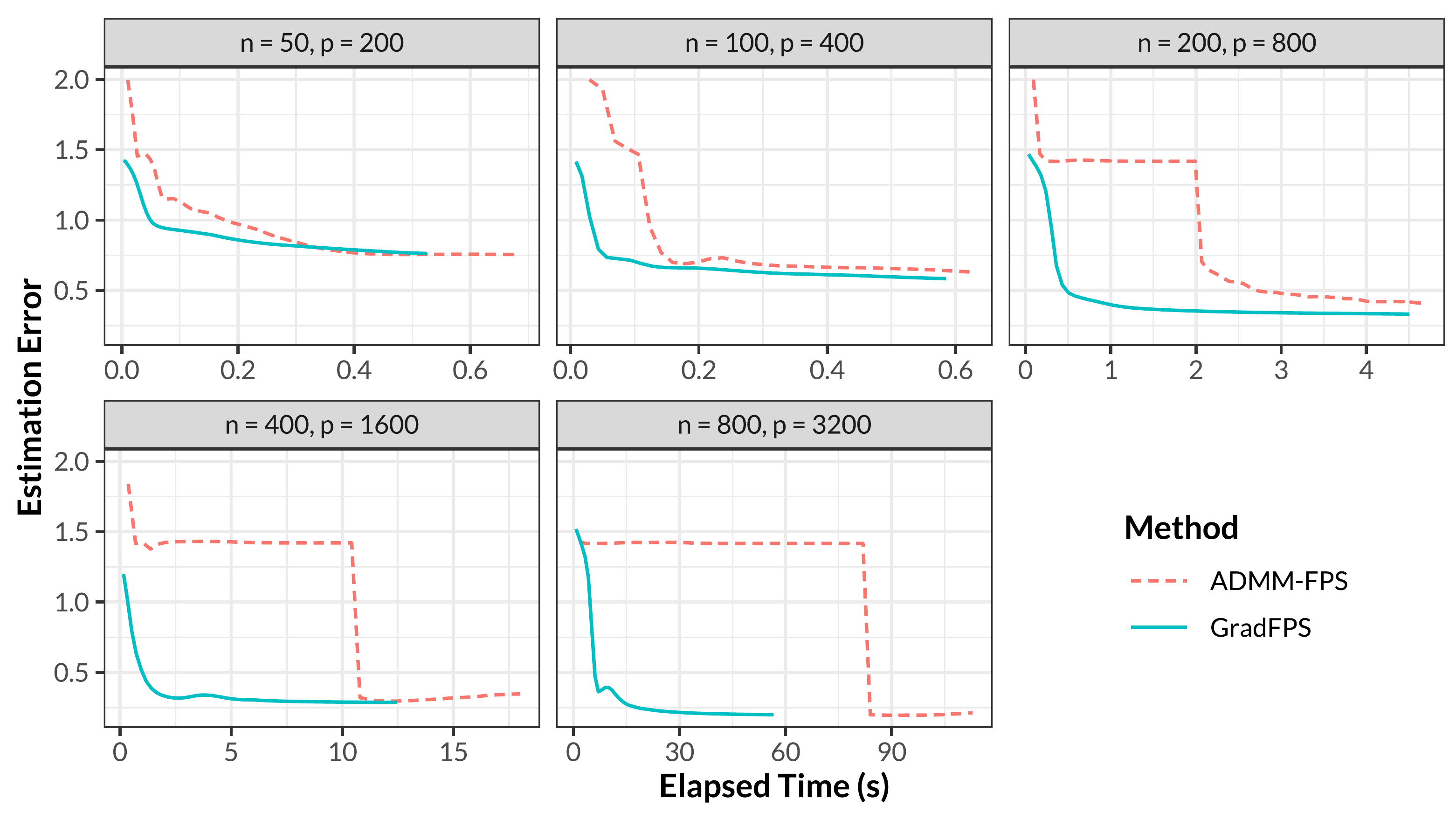}
\par\end{centering}
\caption{\label{fig:computational_efficiency}Comparing the computational efficiency
of the existing ADMM-based algorithm and the proposed GradFPS. The
horizontal axis is the elapsed time in seconds, and the vertical axis
stands for the estimation error $\Vert\hat{X}-\Pi\Vert_{F}$.}
\end{figure}

The first experiment compares the computational efficiency of the
existing ADMM-based algorithm (ADMM-FPS, \citealp{vu2013fantope})
and the proposed GradFPS (Algorithm \ref{alg:grad_fps}) with different
sizes of data. Under each pair of $(n,p)$, a data set $Z_{1},\ldots,Z_{n}$
is simulated to compute the sample covariance matrix $S=n^{-1}\sum_{i=1}^{n}Z_{i}Z_{i}^{\mathrm{T}}$,
and the sparsity parameter is set to $\lambda=0.5\sqrt{\log(p)/n}$.
We compute the estimator $\hat{X}$ using both algorithms with initial
value $X_{0}=V_{2}V_{2}^{\mathrm{T}}$, where $V_{2}$ contains the
top two eigenvectors of $S$. For both algorithms, the best step size
parameter is chosen by trying ten equally-spaced values ranging from
0.01 to 0.1. We then plot the estimation error in each iteration against
the computing time, with the comparison results illustrated in Figure
\ref{fig:computational_efficiency}.

Figure \ref{fig:computational_efficiency} shows a number of interesting
findings. First, as expected, GradFPS has demonstrated superior computational
efficiency compared with ADMM-FPS. It is clear that the curves for
GradFPS decrease very quickly at early stages of the optimization,
which indicates that GradFPS is able to provide reasonably accurate
solutions in a short time. Such a property is crucial, since a common
practice for computing sparse PCA is to use convex solutions as good
initial values for fast nonconvex methods \citep{wang2014nonconvex,chen2015fast,tan2018sparse}.
Second, the curves for ADMM-FPS have irregular shapes, containing
some long ``plateaus'' and even increasing parts. In practice, such
patterns are misleading for convergence tests. In contrast, the curves
for GradFPS mostly show a monotone progress. Finally, even if the
same initial value is supplied to both algorithms, the GradFPS algorithm
tends to make better use of it, as the initial errors of GradFPS are
smaller than those of ADMM-FPS.

\subsection{Online Algorithms}

\begin{figure}
\begin{centering}
\subfloat[]{\includegraphics[width=0.49\textwidth]{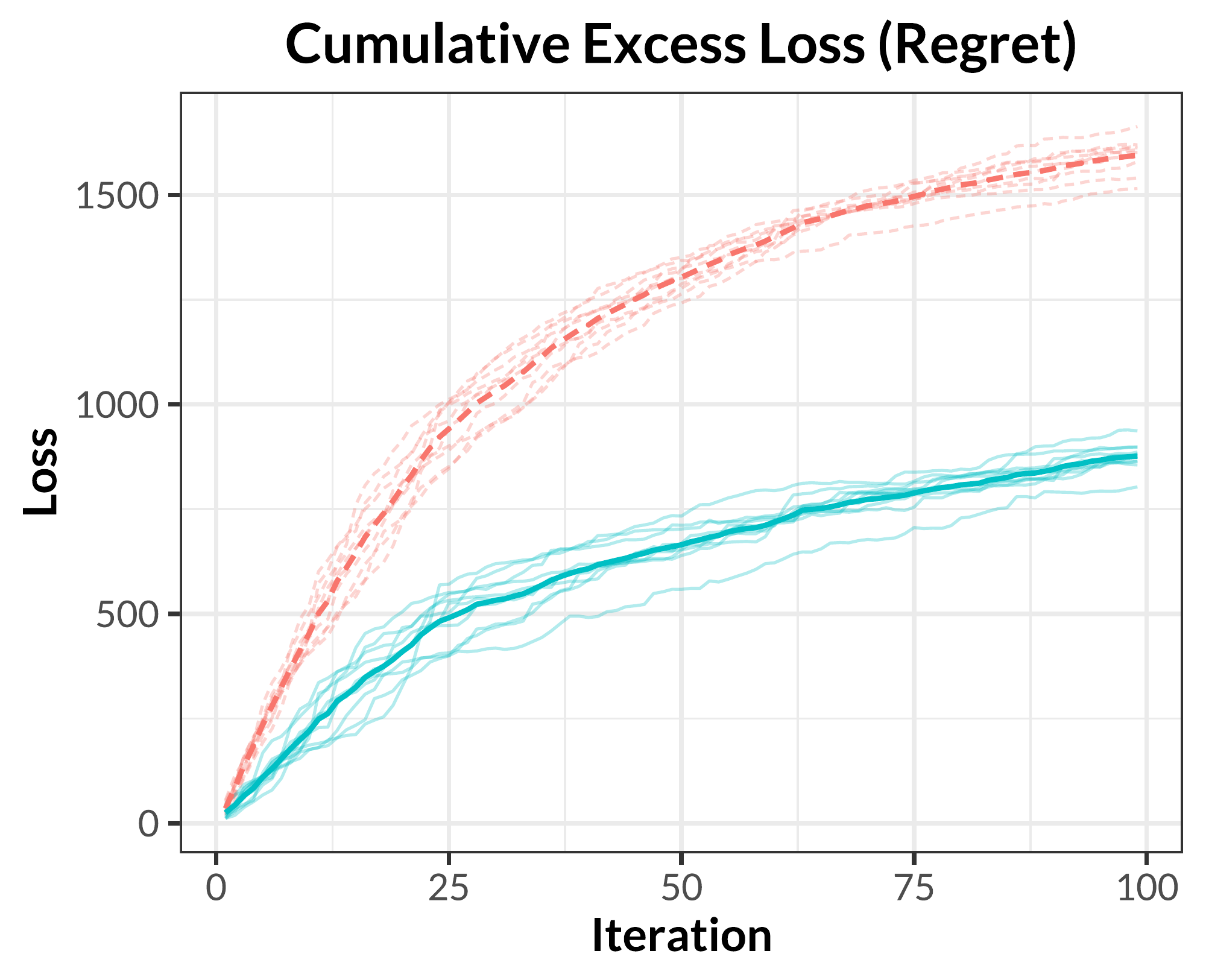}

} \subfloat[]{\includegraphics[width=0.49\textwidth]{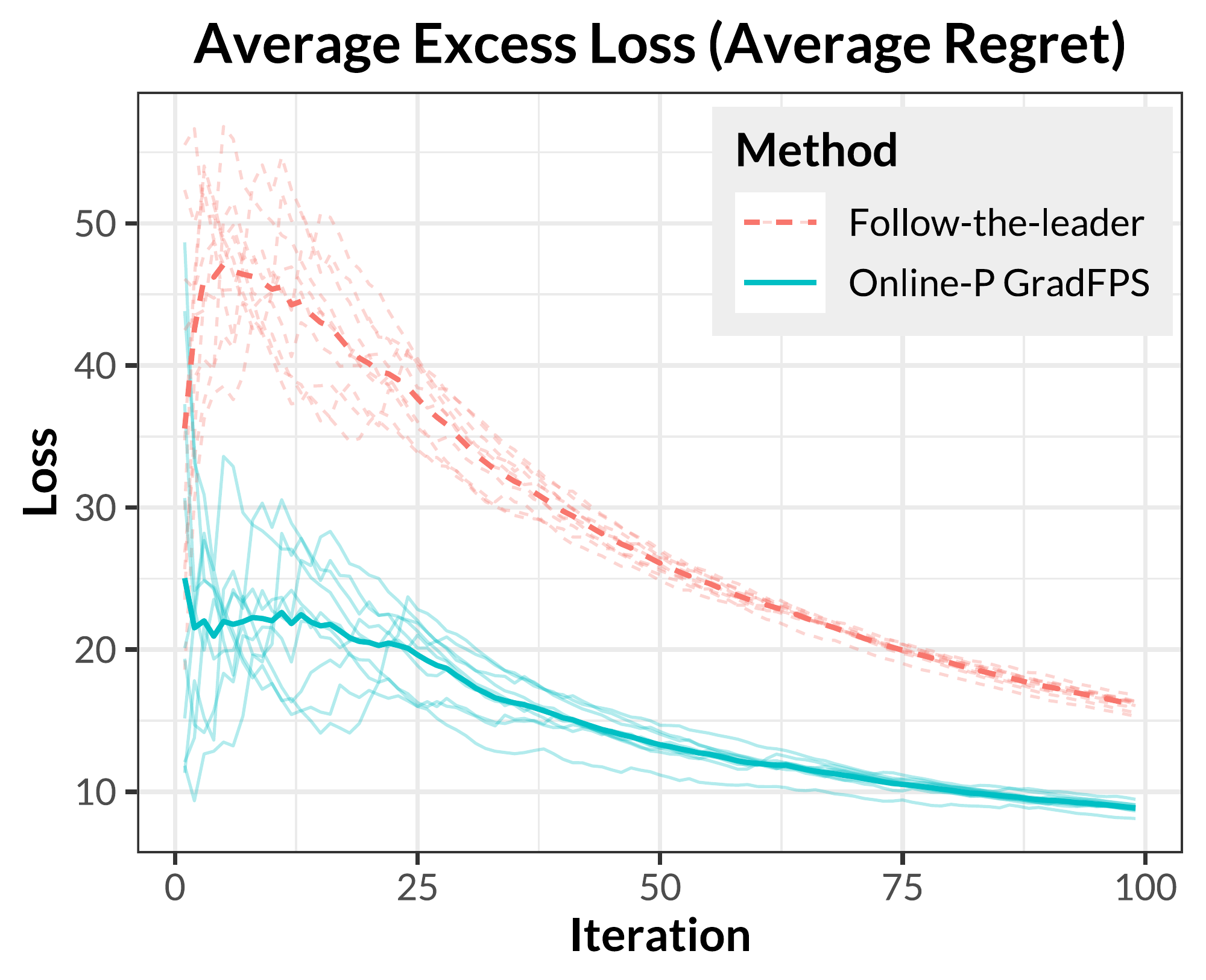}

}
\par\end{centering}
\caption{\label{fig:online_regret}(a) The cumulative regret values for the
two online algorithms. (b) The average regret values. Thin curves
represent ten replications of the experiment, and thick ones stand
for the mean value across experiments.}
\end{figure}

The next experiment studies the behavior of Online-P GradFPS (Algorithm
\ref{alg:mirror_online_fps}) for online sparse PCA, compared with
the naive FTL algorithm. In this case $T=100$, $p=400$, and data
points $Z_{1},\ldots,Z_{T+1}\overset{iid}{\sim}N(0,\Sigma)$ are drawn
in a streaming fashion. We apply the FTL method and Online-P GradFPS
on this data sequence, and compute their regret values at each time
point. To account for the variability in the data generation process,
this experiment is repeated ten times, and Figure \ref{fig:online_regret}(a)
and Figure \ref{fig:online_regret}(b) show the cumulative and average
regret values for the two online algorithms, respectively.

It is clear from Figure \ref{fig:online_regret} that Online-P GradFPS
has much smaller regret values compared with the naive FTL method.
In fact, at the final time point Online-P GradFPS only has about half
of the regret value of FTL. This result implies that the proposed
method is effective in controlling the procedural loss.

\subsection{Comparison between Online-T and Online-P Algorithms}

\begin{figure}
\begin{centering}
\subfloat[]{\includegraphics[width=0.49\textwidth]{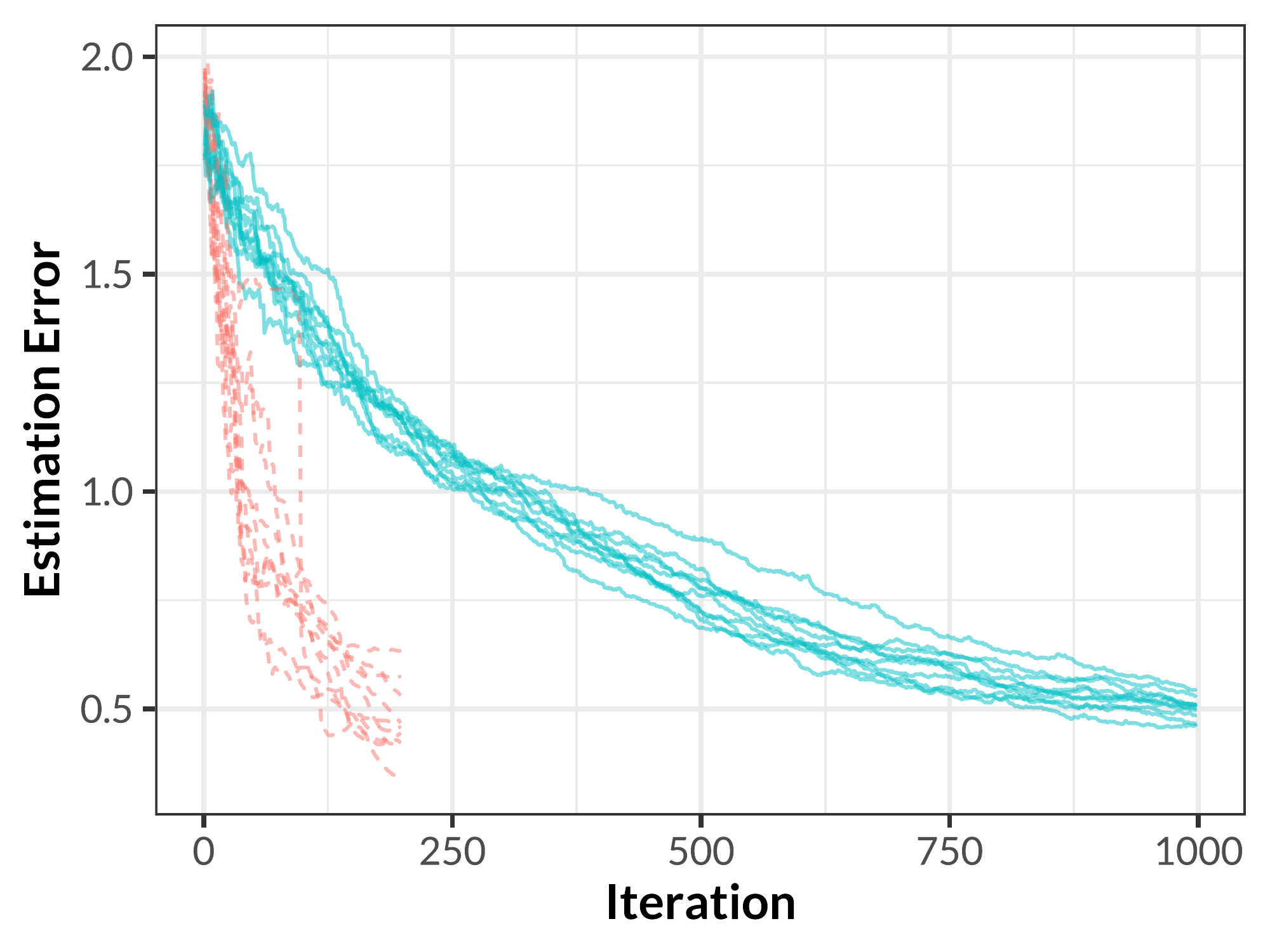}

} \subfloat[]{\includegraphics[width=0.49\textwidth]{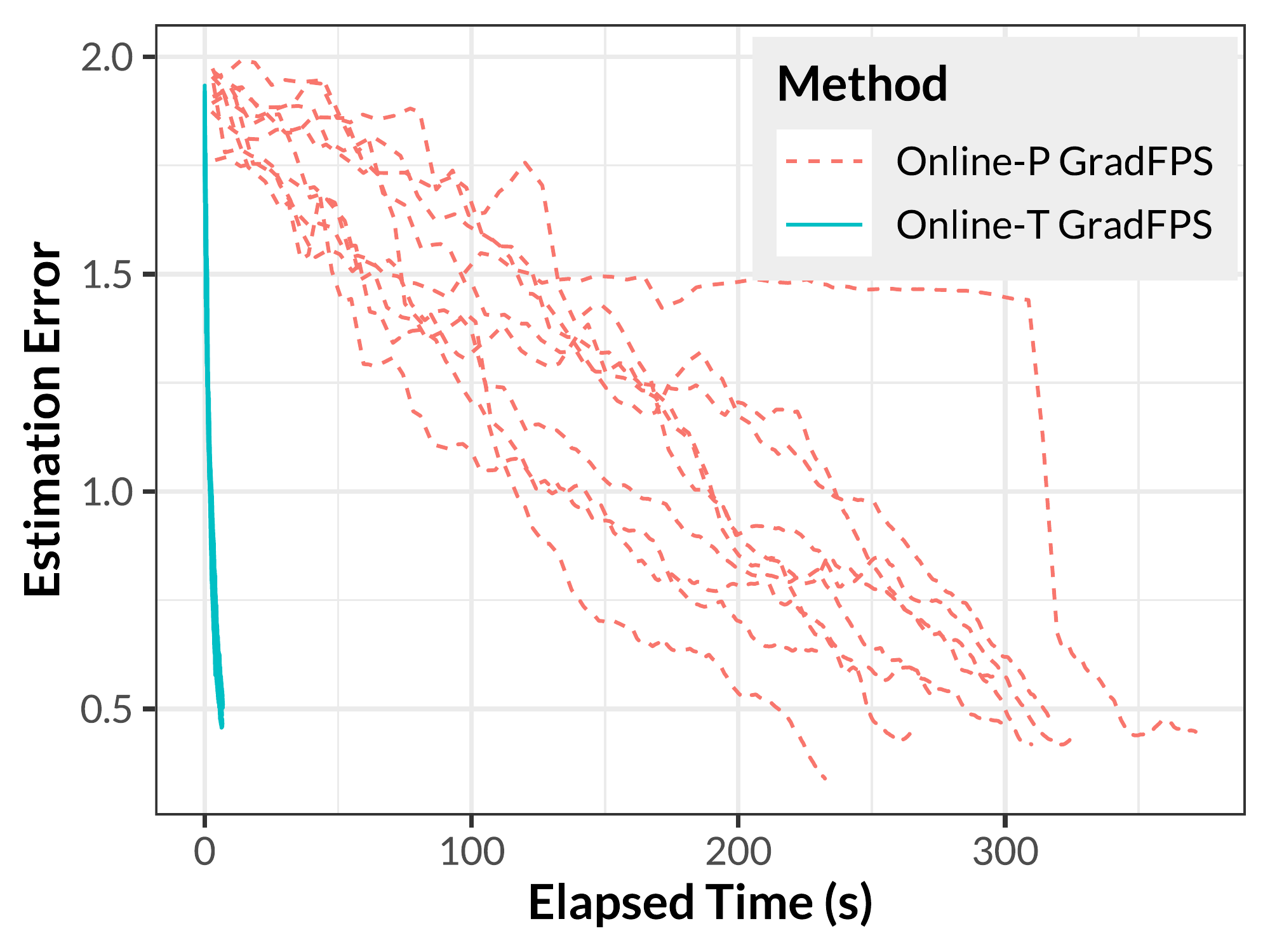}

}
\par\end{centering}
\caption{\label{fig:two_online_algorithms}(a) Plotting the estimation error
$\Vert X_{t}-\Pi\Vert_{F}$ against the iteration index $t$. (b)
The error versus the computing time. Each curve stands for one simulation
run.}
\end{figure}

In Section \ref{sec:online_sparse_pca} we have developed two different
online sparse PCA algorithms, so a natural question is how they compare
to each other. To answer this, we fix $p=200$ and simulate ten streaming
data sets using the model in Section \ref{subsec:simulation_setting}.
Both the Online-T and Online-P GradFPS algorithms are applied to the
data sets, with the former stopped after $T_{1}=1000$ iterations,
and the latter stopped at $T_{2}=200$. The estimation error at each
iteration for both algorithms are shown in Figure \ref{fig:two_online_algorithms}(a).

It is clear that the convergence of Online-P GradFPS is much faster
than Online-T GradFPS in terms of the number of iterations, which
is consistent with the theory developed in Section \ref{sec:online_sparse_pca}.
However, if the $x$-axis is set to the computing time, as illustrated
in Figure \ref{fig:two_online_algorithms}(b), then we find that Online-T
GradFPS is an order of magnitude faster. This phenomenon suggests
the following guideline for choosing the online algorithm: if the
number of data points are limited and the statistical accuracy is
a concern, then Online-P GradFPS is preferred; otherwise, if data
are abundant and computation needs to be fast, then Online-T GradFPS
would be a proper choice.

\section{Application}

In this section we apply sparse PCA to an RNA sequencing data set
to analyze the co-expression relationship among genes. The aim of
our analysis is to detect groups of genes, typically referred to as
modules, with high co-expression. Such an analysis is motivated by
the biological conjecture that genes in the same module are likely
to be functionally related \citep{stuart2003gene}. Sparse PCA is
well suited to this challenging problem for which expression data
are available for tens of thousands of genes.

We study the brain gene expression data collected by the CommonMind
Consortium (CMC), which contain $p=16,423$ genes from 258 schizophrenia
(SCZ) subjects and 279 control subjects \citep{fromer2016gene}. The
control group is used as a baseline, and our main interest is in the
SCZ group. We compute Pearson\textquoteright s correlation coefficients
between genes utilizing the processed and normalized expression data
provided by the CMC, and then apply sparse PCA to the sample correlation
matrix. The number of sparse principal components is chosen to be
$d=5$, and the sparsity parameter $\lambda$ is selected in the following
way. First, we compute the solution paths of sparse PCA in both the
SCZ group and the control group based on a common sequence of $\lambda$
values. Then for each $\lambda$, two active sets $\Omega_{ctr}^{\lambda},\Omega_{scz}^{\lambda}\subset\{1,2,\ldots,p\}$
are determined, where $i\in\Omega_{ctr}^{\lambda}$ if the $i$-th
gene has at least one nonzero factor loading in the five sparse principal
components, and $i\in\Omega_{scz}^{\lambda}$ is defined likewise.
We limit the range of $\lambda$ so that $\min\{|\Omega_{ctr}^{\lambda}|,|\Omega_{scz}^{\lambda}|\}\ge50$
and $\max\{|\Omega_{ctr}^{\lambda}|,|\Omega_{scz}^{\lambda}|\}\le300$,
where $|\Omega|$ denotes the cardinality of a set $\Omega$. Define
the overlapping coefficient as $V(\lambda)=|\Omega_{ctr}^{\lambda}\cap\Omega_{scz}^{\lambda}|/|\Omega_{ctr}^{\lambda}\cup\Omega_{scz}^{\lambda}|$,
and $\lambda$ is chosen to maximize $V(\lambda)$, indicating that
these two groups share maximal common structures. Using this approach,
we finally select $\lambda=0.85$, under which $|\Omega_{ctr}|=292$,
$|\Omega_{scz}|=185$, and $|\Omega_{ctr}\cap\Omega_{scz}|=114$.

\begin{figure}[h]
\begin{centering}
\includegraphics[width=0.45\textwidth]{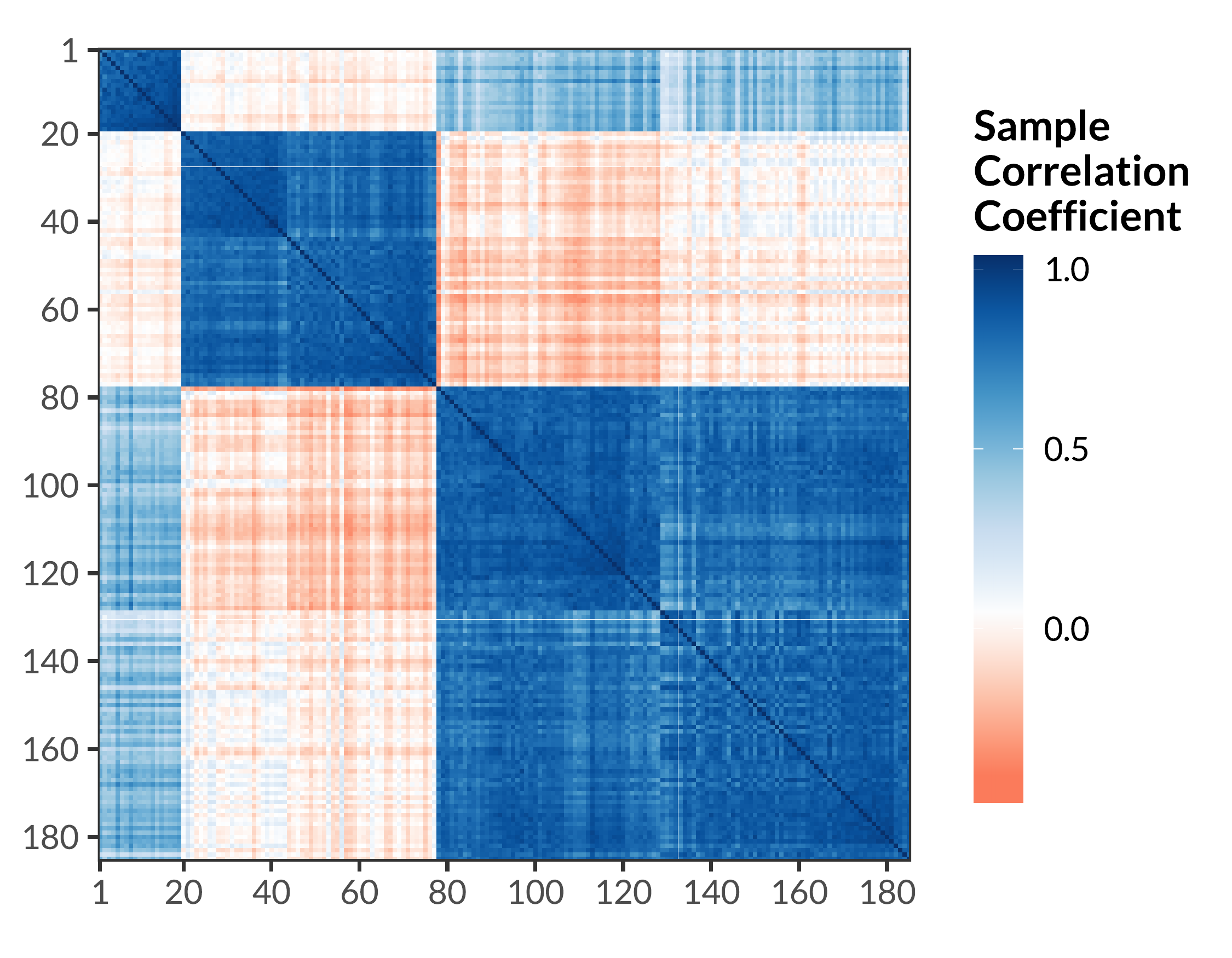}\includegraphics[width=0.55\textwidth]{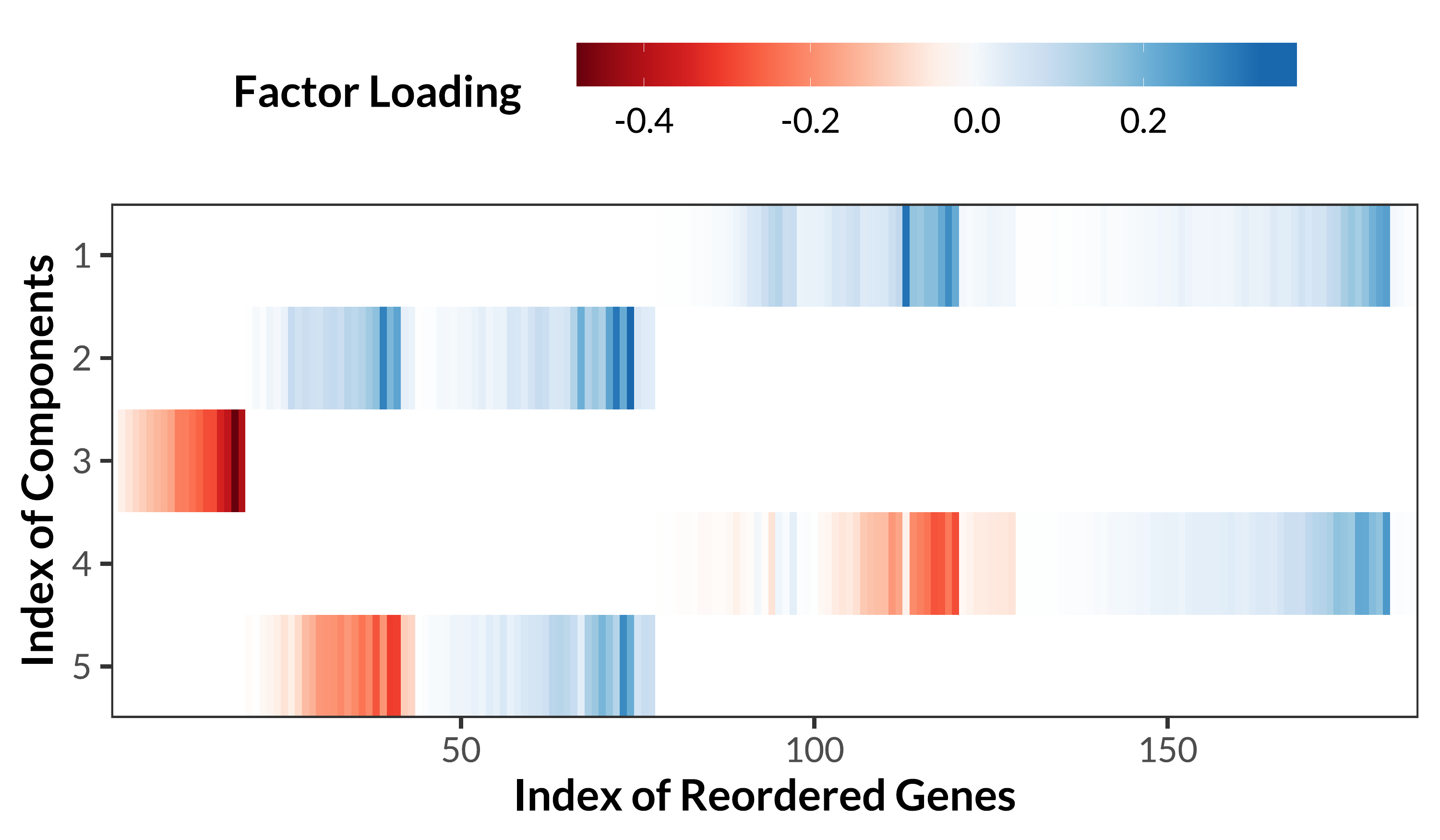}
\par\end{centering}
\caption{\label{fig:reordered_correlation_factor_loading}The reordered sample
correlation matrix of the selected genes in the SCZ group (left) and
the reordered factor loadings (right).}
\end{figure}

After computing the sparse PCA solution for the SCZ group at the selected
$\lambda$, the genes in the active set are clustered based on their
factor loadings, with the number of clusters set to $k=5$. For display,
the indices of genes are reordered so that the genes in the same cluster
are adjacent. Figure \ref{fig:reordered_correlation_factor_loading}
shows the sample correlation matrix and factor loadings based on the
reordered indices of selected genes.

It can be easily observed from Figure \ref{fig:reordered_correlation_factor_loading}
that there are three major modules in the correlation matrix, and
the second and third modules have two sub-modules, respectively, resulting
in five clusters in total. Such a structure is clearly reflected in
the factor loadings, in which the first three components define the
major modules, whereas the last two components add sub-structure to
the second and third modules.

To validate our results, we compare the clusters reflected in Figure
\ref{fig:reordered_correlation_factor_loading} with the modules obtained
by the weighted gene co-expression network analysis (WGCNA, \citealp{zhang2005general}).
Table \ref{tab:cross_tab} demonstrates the cross table for the two
methods of module assignment on the selected genes, where the numbered
modules are given by our approach, and the ones labeled by color names
are the WGCNA results provided by \citet{fromer2016gene}. It is clear
that our modules are well aligned with the WGCNA ones, with three
extra advantages. First, our clusters have smaller sizes and stronger
within-group correlation. For instance, the Green WGCNA module contains
414 genes, whereas our M-1, a subset of the Green module, has only
19 genes. In many studies, researchers are more interested in a small
number of genes that are representative for the whole module. Second,
we have detected highly correlated genes that are assigned to different
modules by WGCNA. As an example, the two genes in the Tan module are
highly correlated with other M-4 genes (a subset of Turquoise), with
average sample correlation coefficients 0.817 and 0.794, respectively.
Finally, our clusters have revealed sub-structure within large modules,
for example M-2 and M-3 are sub-modules for Brown.

\begin{table}[h]
\caption{\label{tab:cross_tab}Cross table for sparse-PCA-based modules (row)
and the WGCNA modules (column). The numbers in the parentheses stand
for the sizes of WGCNA modules.}

\centering{}%
\begin{tabular}{cccccc}
\toprule 
 & Green (414) & Brown (528) & Turquoise (1155) & Tan (248) & Blue (609)\tabularnewline
\midrule
\midrule 
M-1 & 19 & 0 & 0 & 0 & 0\tabularnewline
\midrule 
M-2 & 0 & 24 & 0 & 0 & 0\tabularnewline
\midrule 
M-3 & 0 & 34 & 0 & 0 & 0\tabularnewline
\midrule 
M-4 & 0 & 0 & 49 & 2 & 0\tabularnewline
\midrule 
M-5 & 0 & 0 & 53 & 0 & 4\tabularnewline
\bottomrule
\end{tabular}
\end{table}

\begin{figure}[h]
\begin{centering}
\includegraphics[width=0.333\textwidth]{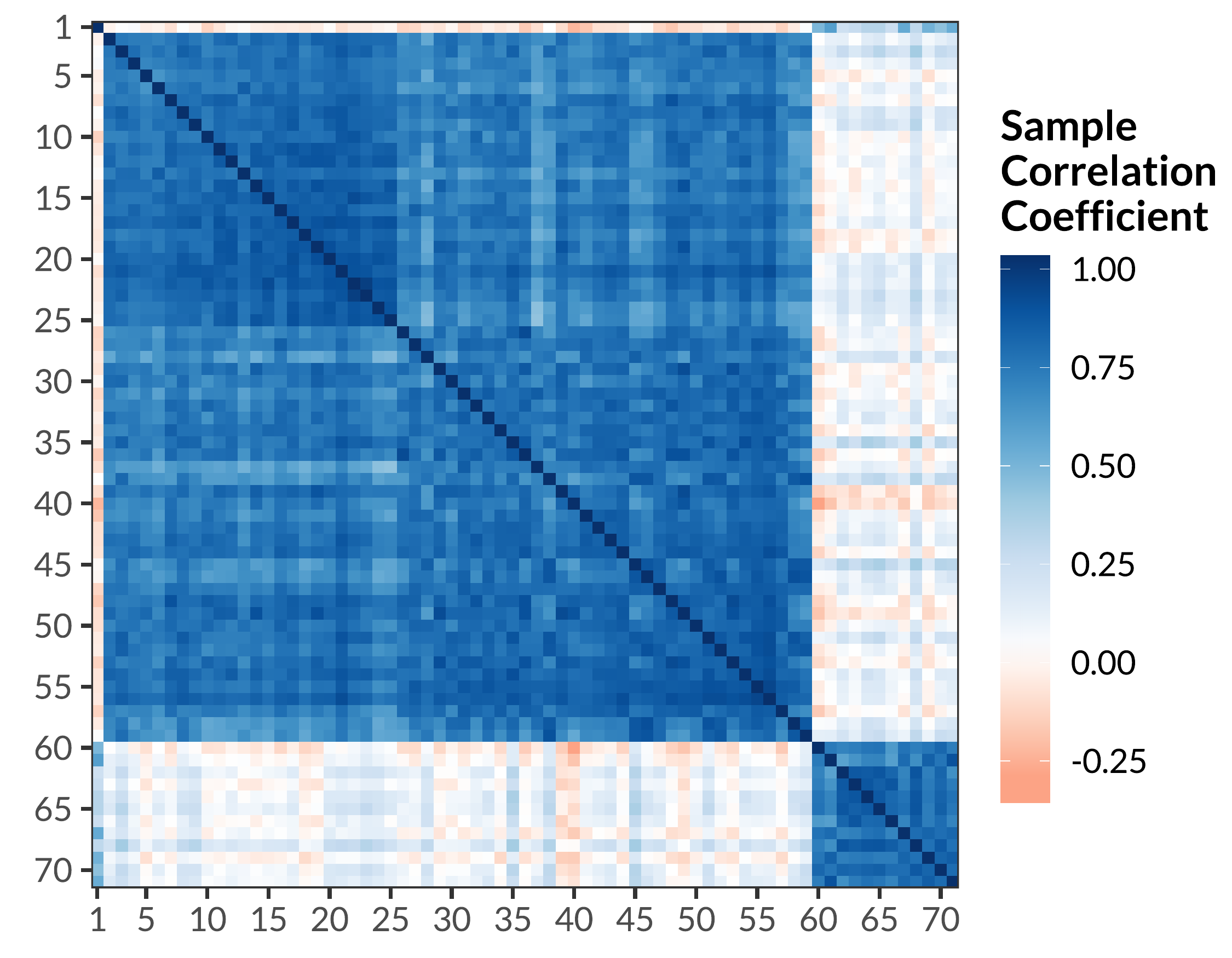}\includegraphics[width=0.333\textwidth]{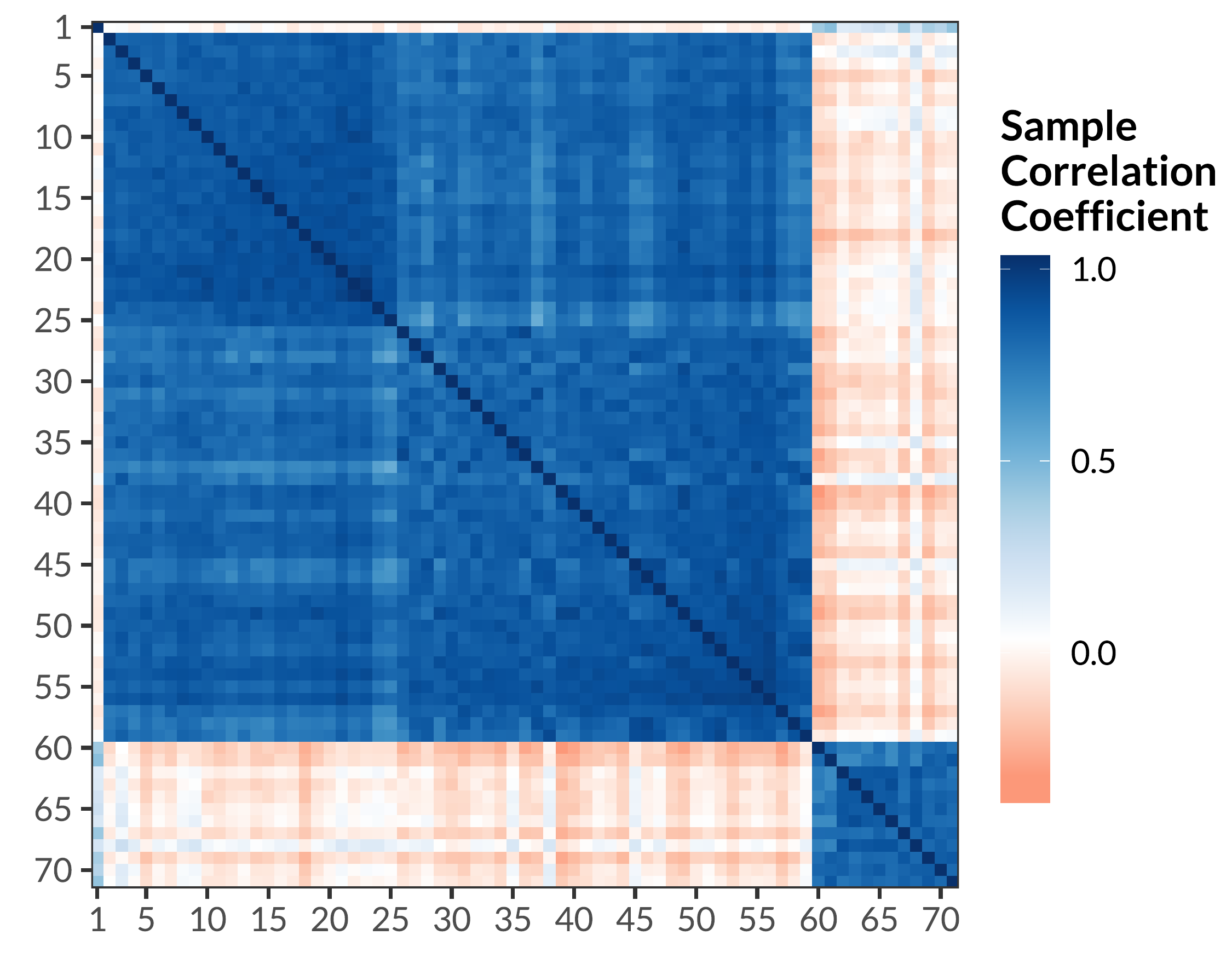}\includegraphics[width=0.333\textwidth]{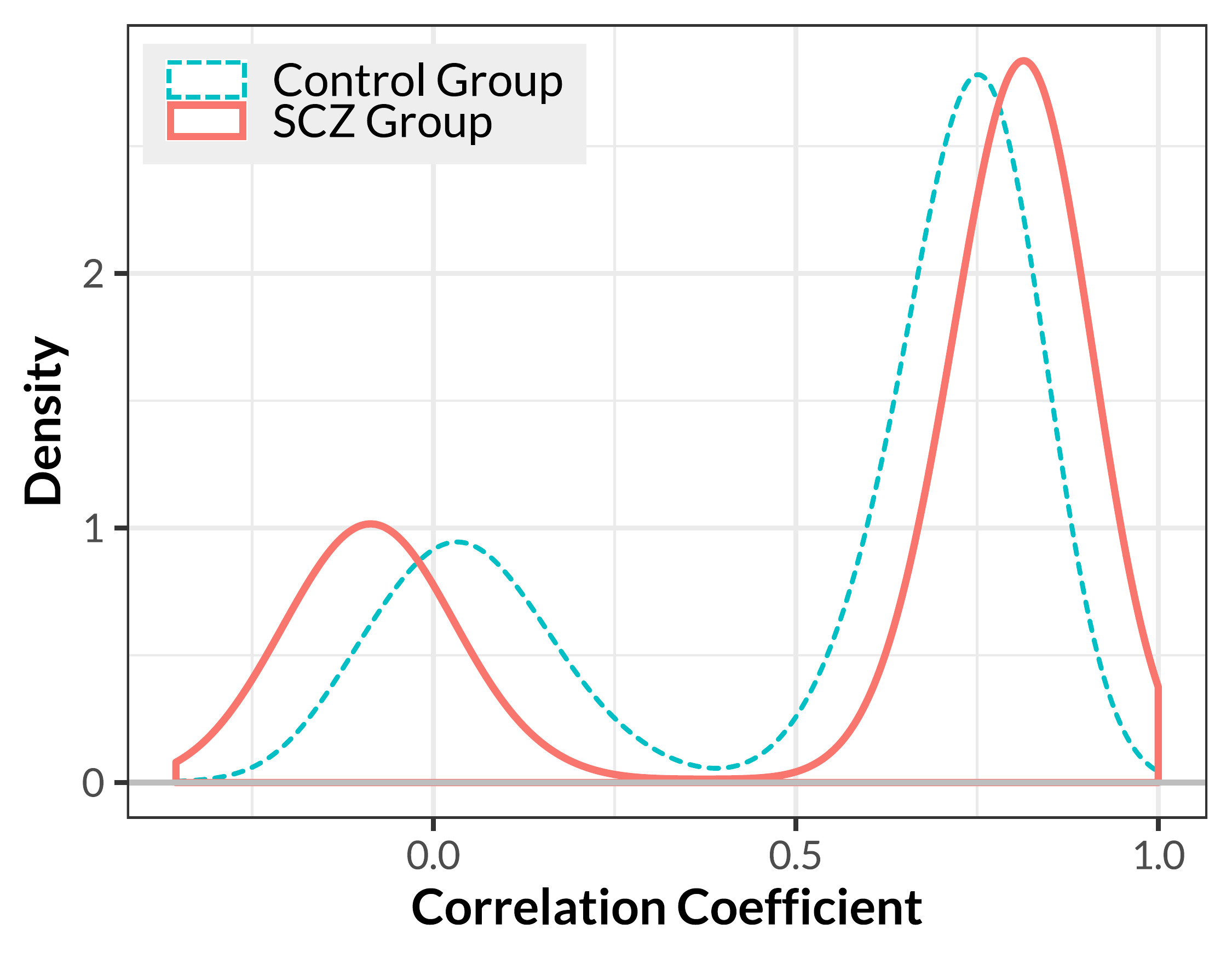}
\par\end{centering}
\caption{\label{fig:compare_correlation_matrices}Comparison of correlation
matrices on SCZ-group-specific genes $\Omega_{scz}^{U}$. Left: the
correlation matrix on $\Omega_{scz}^{U}$ for the control group. Middle:
the correlation matrix for the SCZ group. Right: density curves for
the off-diagonal correlation coefficients.}
\end{figure}

Next, by comparing with the control group, we study the structural
change of gene co-expression relationship in the SCZ group. Consider
the genes that are selected in the SCZ group but not in the control
group, forming the gene set $\Omega_{scz}^{U}=\Omega_{scz}\backslash\Omega_{ctr}$.
Figure \ref{fig:compare_correlation_matrices} illustrates the sample
correlation matrices on $\Omega_{scz}^{U}$ for both the control group
(left panel) and the SCZ group (middle panel). In addition, to better
visualize the correlation pattern, density curves of off-diagonal
correlation coefficients are shown in the right panel of Figure \ref{fig:compare_correlation_matrices}.

Figure \ref{fig:compare_correlation_matrices} highlights an interesting
difference between the control group and the SCZ group. In both groups,
the correlation matrices indicate a similar two-block structure, but
density curves of the correlations summarize the differences between
groups. Both exhibit two modes, representing the between-module and
within-module correlation coefficients, respectively; however, the
coefficients in the SCZ group are obviously more extreme than those
in the control group. The first mode differs in sign, indicating that
the small positive between-module correlations in the control group
are largely negative in the SCZ group. These findings provide insights
for future studies of schizophrenia based on brain gene expression
data.

\section{Conclusion and Discussion}

In this article we have developed a novel efficient algorithm for
the convex sparse PCA model, which is shown to outperform the existing
ADMM-based method in many aspects. The main technique used is to transform
the original highly constrained optimization problem into an unconstrained
one, so that gradient-based and projection-free algorithms can be
applied to seek the solution. This technique also allows us to compute
sparse PCA for large-scale streaming data, leading to various online
learning algorithms.

We point out that this framework of analysis has a great potential
for further extensions, and below we mention two possible future directions
for research. First, within the sparse PCA framework, the efficient
algorithm can be developed for other types of problems that come with
a different penalty term, such as the trend filtering \citep{tibshirani2014adaptive}
or the localized functional PCA \citep{chen2015localized}. Other
types of penalty terms are also applicable as long as they are convex
functions. Second, the two technical tools developed in this article,
namely the gradient-based and projection-free optimization method
for highly constrained problems, and the analysis of online learning
algorithms, can be extended to other interesting statistical models.
An example of this kind is the graphical lasso \citep{friedman2008sparse},
in which the precision matrix is constrained in the positive semidefinite
cone with an elementwise $\ell_{1}$ penalty. Similar to sparse PCA,
online learning algorithms may be developed for graphical lasso using
an unconstrained formulation of the objective function.

\section*{Acknowledgments}

This work was supported by NIMH grants R37MH057881-22 and R37MH057881-22S,
and NSF grant DMS-1553884.

Data were generated as part of the CommonMind Consortium supported
by funding from Takeda Pharmaceuticals Company Limited, F. Hoffman-La
Roche Ltd and NIH grants R01MH085542, R01MH093725, P50MH066392, P50MH080405,
R01MH097276, RO1-MH-075916, P50M096891, P50MH084053S1, R37MH057881,
AG02219, AG05138, MH06692, R01MH110921, R01MH109677, R01MH109897,
U01MH103392, and contract HHSN271201300031C through IRP NIMH. Brain
tissue for the study was obtained from the following brain bank collections:
the Mount Sinai NIH Brain and Tissue Repository, the University of
Pennsylvania Alzheimer\textquoteright s Disease Core Center, the University
of Pittsburgh NeuroBioBank and Brain and Tissue Repositories, and
the NIMH Human Brain Collection Core. CMC Leadership: Panos Roussos,
Joseph Buxbaum, Andrew Chess, Schahram Akbarian, Vahram Haroutunian
(Icahn School of Medicine at Mount Sinai), Bernie Devlin, David Lewis
(University of Pittsburgh), Raquel Gur, Chang-Gyu Hahn (University
of Pennsylvania), Enrico Domenici (University of Trento), Mette A.
Peters, Solveig Sieberts (Sage Bionetworks), Thomas Lehner, Stefano
Marenco, Barbara K. Lipska (NIMH).

\appendix
\addtolength{\jot}{-0.3em}

\section{Appendix}

\subsection{Expressions for constants and bounds}

\label{subsec:constants}

\noindent \textbf{Theorem \ref{thm:batch_fps_convergence}}: The
constant is $C=\max\{\alpha^{-1}(C_{0}^{2}+4C_{0}L_{g}),2C_{0}L_{g}\}+2C_{0}L_{g}$,
where
\[
L_{g}=\sqrt{(\lambda p)^{2}+\{\Vert S\Vert_{F}+\mu(1+\sqrt{(p+d)(d+1)})\}^{2}},
\]
and $C_{0}>0$ is a constant that only depends on $X_{0}$ and the
optimal point of the optimization problem.

\noindent \textbf{Theorem \ref{thm:prox_online_fps_convergence}}:
The regret bound in explicit form is given by
\begin{equation}
\frac{1}{T}\mathcal{R}(\{X_{t}\},T)\le\frac{2d/\alpha_{0}+\alpha_{0}C_{2}}{\sqrt{T}}+\frac{\alpha_{0}}{2T}\sum_{t=1}^{T}\frac{\Vert S_{t+1}\Vert_{F}^{2}+C_{1}\Vert S_{t+1}\Vert_{F}}{\sqrt{t}},\label{eq:prox_online_fps_regret}
\end{equation}
and the estimation error bound is $\Vert\hat{X}_{T}-\Pi\Vert_{F}\le C(T)+\sqrt{2/\delta_{d}}\cdot\sqrt{C(T)+\lambda s\sqrt{d}}$,
where $C(T)=C_{3}/\sqrt{T}+C_{4}\{\log(T)+1\}/T=\mathcal{O}(1/\sqrt{T})$.
The relevant constants are
\begin{align*}
C_{1} & =\lambda p+\nu\sqrt{p(p+d)}+\nu\sqrt{p/(d+1)},\\
C_{2} & =\nu^{2}p(p+d)+2(\lambda p)^{2}+2\lambda p\nu\sqrt{p(p+d)}+2\nu\sqrt{p/(d+1)}C_{1},\\
C_{3} & =2d/\alpha_{0}+D_{1}+\alpha_{0}\{C_{2}+C_{1}(\mu_{1}+\Vert\Sigma\Vert_{F})+\mu_{2}\},\\
C_{4} & =\alpha_{0}(C_{1}D_{2}+D_{3})/2,
\end{align*}
where $D_{1}=\max\left\{ 2b_{1}\varepsilon_{l},2\sigma_{1}\sqrt{2d\varepsilon_{l}}\right\} $,
$D_{2}=\max\left\{ 2b_{1}\varepsilon_{l},\sigma_{1}\sqrt{2\varepsilon_{l}}\right\} $,
$D_{3}=\max\left\{ 2b_{2}\varepsilon_{l},\sigma_{2}\sqrt{2\varepsilon_{l}}\right\} $,
and $\varepsilon_{l}=\log(3/\varepsilon)$.

\noindent \textbf{Theorem \ref{thm:mirror_online_fps_convergence}}:
The regret bound in explicit form is given by
\[
\frac{1}{T}\mathcal{R}(\{X_{t}\},T)\le\frac{\Vert\Pi\Vert_{r,r}^{2}}{2\sqrt{T}}+\frac{1}{T}\sum_{t=1}^{T}\left\{ (\psi_{t+1}+\lambda)\varepsilon_{t}+\frac{\beta\nu\sqrt{t}\varepsilon_{t}^{2}}{2}\right\} +\frac{1}{2\beta T}\sum_{t=1}^{T}\frac{\psi_{t+1}^{2}}{\sqrt{t}},
\]
where $\psi_{t}=\Vert S_{t}\Vert_{\infty,\infty}$. The bound for
the estimation error is
\[
\Vert\hat{X}_{T}-\Pi\Vert_{F}\le\frac{4\sigma s\sqrt{\log(p)}+2s^{2-4/\log(p)}\sqrt{d}}{\delta_{d}\sqrt{T}+\beta}+\frac{\beta}{2\sqrt{T}}+\frac{\sqrt{\beta/(\delta_{d}+\beta/\sqrt{T})}}{T^{3/4}}.
\]

\subsection{Computation of $\mathbf{prox}_{\alpha f_{2}}(X)$}

\label{subsec:computation_prox_f2}

By definition $\mathbf{prox}_{\alpha f_{2}}(X)=\arg\min_{U\in\mathcal{X}}\left\{ f_{2}(U)+(2\alpha)^{-1}\Vert U-X\Vert_{F}^{2}\right\} $,
so an easy iterative method has the form $U_{k+1}=\mathcal{P}_{\mathcal{X}}(U_{k}-\alpha\eta_{k}\nabla f_{2}(U_{k})-\eta_{k}(U_{k}-X))$,
where $\eta_{k}$ is the step size. Since the objective function is
strongly convex, this method converges at the speed of $\mathcal{O}(1/K)$,
where $K$ is the number of iterations.

The direct method for computing $\mathbf{prox}_{\alpha f_{2}}(X)$
is based on the following observation. Let $\theta_{1}\ge\cdots\ge\theta_{p}$
be the eigenvalues of $X$, and $\gamma_{1},\ldots,\gamma_{p}$ be
the associated eigenvectors. If $\mu$ is sufficiently large, then
$\mathbf{prox}_{\alpha f_{2}}(X)=\sum_{i=1}^{p}u_{i}\gamma_{i}\gamma_{i}^{\mathrm{T}}$,
where
\begin{equation}
u=(u_{1},\ldots,u_{p})^{\mathrm{T}}=\underset{\substack{u_{1}+\cdots+u_{p}=d\\
0\le u_{i}\le1
}
}{\arg\min}\,\sum_{i=1}^{p}\left\{ -\theta_{i}u_{i}+\frac{1}{2\alpha}(\theta_{i}-u_{i})^{2}\right\} \label{eq:quadratic_programming}
\end{equation}
is the solution to a quadratic programming problem. Most importantly,
the elements in $u$ has a decreasing order, $u_{1}\ge\cdots\ge u_{p}$,
and for some index $t$ we have $u_{i}=0$ for $i\ge t$. Therefore,
we can sequentially compute the eigenvalues $\theta_{1}\ge\cdots\ge\theta_{t}$
until $u_{t}=0$ is met. In this way the full decomposition of $X$
is avoided.

\subsection{Solving the subproblem of Algorithm \ref{alg:mirror_online_fps}}

\label{subsec:subproblem_omd}

Denote $S_{1:t}=t^{-1}Y_{t}=t^{-1}\sum_{i=1}^{t}S_{i}$ and $\mu_{t}=(\Vert S_{1:t}\Vert_{F}+\lambda p+\exp(-4)\sqrt{d/t}p^{2}+1)\sqrt{p/(d+1)}$,
and then we have $\min_{X\in\mathcal{X}}\mathring{\mathcal{L}}(X;Y_{t},t)=\min_{X\in\mathcal{X}}\,\{f_{1}(X)+f_{2}(X)+f_{3}(X)\}$,
where $f_{1}(X)=\lambda\Vert X\Vert_{1}$, $f_{2}(X)=-\mathrm{tr}(S_{1:t}X)+\mu_{t}\left(d_{C_{1}}(X)+r_{1}[g_{1}(X)]_{+}+r_{2}[g_{2}(X)]_{+}\right)$,
and $f_{3}(X)=\Vert X\Vert_{r,r}^{2}/(2\sqrt{t})$. The subproblem
of Algorithm \ref{alg:mirror_online_fps} can be solved using the
procedure in Algorithm \ref{alg:sub_problem}.

\begin{algorithm}[h]
\caption{\label{alg:sub_problem}Solving $\min_{X\in\mathcal{X}}\,\mathring{\mathcal{L}}(X;Y_{t},t)$}

\begin{spacing}{1.5}

\begin{algorithmic}[1]

\REQUIRE $Y_{t}$, $K$, $\alpha$, initial value $X_{0}\in\mathcal{X}$

\ENSURE $\hat{X}$

\STATE $Z_{0}^{(1)}=Z_{0}^{(2)}=Z_{0}^{(3)}\leftarrow X_{0}$

\FOR{ $k=0,1,\ldots,K-1$ }

\STATE $\bar{Z}_{k}\leftarrow(Z_{k}^{(1)}+Z_{k}^{(2)}+Z_{k}^{(3)})/3$

\STATE $X_{k+1}\leftarrow\mathcal{P}_{\mathcal{X}}\left(\bar{Z}_{k}\right)=\min\left\{ 1,\sqrt{d}/\Vert\bar{Z}_{k}\Vert_{F}\right\} \cdot\bar{Z}_{k}$

\STATE $Z_{k+1}^{(1)}\leftarrow Z_{k}^{(1)}-X_{k+1}+\mathbf{prox}_{\alpha f_{1}}(2X_{k+1}-Z_{k}^{(1)})$

\STATE $Z_{k+1}^{(2)}\leftarrow Z_{k}^{(2)}-X_{k+1}+\mathbf{prox}_{\alpha f_{2}}(2X_{k+1}-Z_{k}^{(2)})$

\STATE $Z_{k+1}^{(3)}\leftarrow Z_{k}^{(3)}-X_{k+1}+\mathbf{prox}_{\alpha f_{3}}(2X_{k+1}-Z_{k}^{(3)})$

\ENDFOR

\RETURN $\hat{X}=K^{-1}\sum_{k=1}^{K}X_{k}$

\end{algorithmic}

\end{spacing}
\end{algorithm}

The proximal operator $\mathbf{prox}_{\alpha f_{1}}$ has closed-form
solution $\mathbf{prox}_{\alpha f_{1}}(X)=\mathcal{S}_{\alpha\lambda}(X)$.
The computation for $\mathbf{prox}_{\alpha f_{2}}$ is given in Appendix
\ref{subsec:computation_prox_f2}. The last operator $\mathbf{prox}_{\alpha f_{3}}$
requires solving the problem $\min_{X\in\mathcal{X}}\{\Vert X\Vert_{r,r}^{2}/(2\sqrt{t})+\Vert X-V\Vert_{F}^{2}/(2\alpha)\}$,
which can be accomplished using the coordinate descent method.

\subsection{Proof of Theorem \ref{thm:unconstrained_problem}}

We first prove an important fact: under Assumption \ref{assu:g_cond},
$[g_{i}(x)]_{+}\ge\rho_{i}d_{G_{i}}(x)$ for all $x\in\mathcal{X}$,
$i=1,\ldots,m$. This result was briefly given in \citet{mahdavi2012stochastic}
with a stronger condition that $\mathcal{X}=\mathbb{R}^{p}$, and
below is our formal proof.

If $g_{i}(x)=0$, then $d_{G_{i}}(x)$ is also zero, so the inequality
holds trivially. In what follows we assume that $g_{i}(x)>0$. By
definition, $d_{G_{i}}^{2}(x)=\min_{g_{i}(y)\le0}\,\Vert y-x\Vert^{2}$,
and the Lagrangian for this constrained optimization problem is $l(y,\lambda)=\Vert y-x\Vert^{2}+\lambda g_{i}(y)$,
with the optimality conditions
\begin{align}
g_{i}(y_{*})\le0,\ \lambda_{*} & \ge0,\nonumber \\
\lambda_{*}g_{i}(y_{*}) & =0,\label{eq:opt_cond_3}\\
2(y_{*}-x)+\lambda_{*}\partial g_{i}(y_{*}) & \ni0.\label{eq:opt_cond_4}
\end{align}
Here $y_{*}$ and $\lambda_{*}$ are the primal and dual optimal points,
respectively. By definition, $y_{*}=\mathcal{P}_{G_{i}}(x)$, and
Assumption \ref{assu:g_cond}(a) indicates that $y_{*}\in\mathcal{X}$.
Since we have assumed that $g_{i}(x)>0$, it is easy to see that $y_{*}-x\neq\mathbf{0}$,
and hence $\lambda_{*}\neq0$ by (\ref{eq:opt_cond_4}). Consequently,
$g_{i}(y_{*})=0$ by (\ref{eq:opt_cond_3}).

Let $\nabla g_{i}(y_{*})$ be the subgradient such that $2(y_{*}-x)+\lambda_{*}\nabla g_{i}(y_{*})=\mathbf{0}$,
and then we have $[\nabla g_{i}(y_{*})]^{\mathrm{T}}(x-y_{*})=\Vert x-y_{*}\Vert\cdot\Vert\nabla g_{i}(y_{*})\Vert$.
Since $g_{i}(x)$ is convex, it holds that
\[
g_{i}(x)\ge g_{i}(y_{*})+[\nabla g_{i}(y_{*})]^{\mathrm{T}}(x-y_{*})=\Vert x-y_{*}\Vert\cdot\Vert\nabla g_{i}(y_{*})\Vert\ge\rho_{i}\Vert x-y_{*}\Vert,
\]
where the last inequality is from Assumption \ref{assu:g_cond}(b).
Finally by definition, $d_{G_{i}}(x)=\Vert y_{*}-x\Vert$, so the
desired inequality holds.

Next we prove part (a) of the theorem. The proof is similar to that
of Proposition 2 of \citet{kundu2018convex}, but under our generalized
settings. Since $f(x)$ is Lipschitz continuous on $\mathcal{X}$,
we have $f(y)-f(x)\le L\Vert y-x\Vert$ for all $x,y\in\mathcal{X}$.
Set $y=\mathcal{P}_{\mathcal{K}}(x)$, and then
\begin{equation}
f_{*}\le f(y)\le f(x)+L\Vert y-x\Vert=f(x)+L\cdot d_{\mathcal{K}}(x).\label{eq:f_lipschitz_ineq}
\end{equation}
On one hand, for $\mu\ge\gamma L$ and all $x\in\mathcal{X}$,
\begin{equation}
\mathcal{L}(x;\mu)\ge f(x)+\mu h\left(d_{C_{1}}(x),\ldots,d_{C_{l}}(x),d_{G_{1}}(x),\ldots,d_{G_{m}}(x)\right)\ge f(x)+\frac{\mu}{\gamma}d_{\mathcal{K}}(x)\ge f_{*},\label{eq:L_f_one_side}
\end{equation}
which indicates that $\mathcal{L}_{*}\ge f_{*}$. On the other hand,
$d_{C_{i}}(x)=[g_{i}(x)]_{+}=0$ for all $x\in\mathcal{K}$, so $\mathcal{L}(x;\mu)=f(x)$
on $\mathcal{K}$. Therefore,
\[
\mathcal{L}_{*}=\min_{x\in\mathcal{X}}\,\mathcal{L}(x;\mu)\le\min_{x\in\mathcal{K}}\,\mathcal{L}(x;\mu)=\min_{x\in\mathcal{K}}\,f(x)=f_{*}.
\]
As a result, we must have $\mathcal{L}_{*}=f_{*}$.

For part (b), if $\mu\ge\gamma(L+1)$, then
\[
f(x_{\varepsilon})+\frac{\mu}{\gamma}d_{\mathcal{K}}(x_{\varepsilon})\le_{(i)}\mathcal{L}(x_{\varepsilon};\mu)\le f_{*}+\varepsilon\le_{(ii)}f(x_{\varepsilon})+L\cdot d_{\mathcal{K}}(x_{\varepsilon})+\varepsilon,
\]
where (i) is true by (\ref{eq:L_f_one_side}), and (ii) holds due
to (\ref{eq:f_lipschitz_ineq}). Hence we get $\Vert x_{\varepsilon}-y_{\varepsilon}\Vert=d_{\mathcal{K}}(x_{\varepsilon})\le\varepsilon$.
Finally, using (\ref{eq:f_lipschitz_ineq}) again yields
\[
\mathcal{L}(y_{\varepsilon};\mu)=f(y_{\varepsilon})\le f(x_{\varepsilon})+L\cdot d_{\mathcal{K}}(x_{\varepsilon})\le\mathcal{L}(x_{\varepsilon};\mu)\le\mathcal{L}_{*}+\varepsilon.
\]

\subsection{Proof of Theorem \ref{thm:dist_ineq}}

Define $U=[0,1]^{p}$, $T=\{z\in\mathbb{R}^{p}:z_{1}+\cdots+z_{p}=d\}$,
and $F=U\cap T$. Let $\theta=(\theta_{1},\ldots,\theta_{p})^{\mathrm{T}}$
be the $p$ eigenvalues of $X$, and then $X\in\mathcal{F}_{1}\Leftrightarrow\theta\in T$,
$X\in\mathcal{F}_{2,3}\Leftrightarrow\theta\in U$, and $X\in\mathcal{F}^{d}\Leftrightarrow\theta\in F$.
It is also easy to see that $d_{T}(\theta)=d_{\mathcal{F}_{1}}(X)$,
$d_{U}(\theta)=d_{\mathcal{F}_{2,3}}(X)$, and $d_{F}(\theta)=d_{\mathcal{\mathcal{F}}^{d}}(X)$,
so it suffices to prove the following inequality for any $z\in\mathbb{R}^{p}$:
\[
d_{F}(z)\le\sqrt{p/(d+1)}\cdot d_{T}(z)+\sqrt{p}\cdot d_{U}(z).
\]

For any $x\in\partial F$, the normal cone of $F$ at $x$ is defined
by $N_{F}(x)=\{y:y^{\mathrm{T}}(x-x')\ge0,\forall x'\in F\}$. Below
are three important facts about normal cones:
\begin{enumerate}
\item It holds that
\begin{equation}
y\in N_{F}(x)\Leftrightarrow x\in\underset{x'\in F}{\arg\max}\ y^{\mathrm{T}}x'.\label{eq:defn_normal_cone}
\end{equation}
\item For all $x\in F$, $x=P_{F}(z)$ if and only if $z-x\in N_{F}(x)$.
\item For all $x\in F$, $y\in N_{F}(x)$, and $t\ge0$, $P_{F}(x+ty)=x$.
\end{enumerate}
Our final goal is to show that there exist constants $c_{1}>0$ and
$c_{2}>0$ such that
\begin{equation}
d_{F}(z)\le c_{1}d_{U}(z)+c_{2}d_{T}(z)\label{eq:bounded_distance}
\end{equation}
for any $z\in\mathbb{R}^{p}$. Using the second fact about normal
cones, we can decompose $z$ as $z=x+y$, where $x=P_{F}(z)$ and
$y\in N_{F}(x)$. For $x=(x_{1},\ldots,x_{p})^{\mathrm{T}}\in\partial F$,
we divide it into three blocks with index sets $I_{1}$, $I_{2}$,
and $I_{3}$ such that
\[
\begin{cases}
x_{k}=1, & k\in I_{1},\\
0<x_{k}<1, & k\in I_{2},\\
x_{k}=0, & k\in I_{3}.
\end{cases}
\]
For simplicity, we can assume $I_{1}=\{1,\ldots,i\}$, $I_{2}=\{i+1,\ldots,j\}$,
and $I_{3}=\{j+1,\ldots,p\}$ without loss of generality. Since $x\in F\subset T$,
we have $\sum_{i=1}^{p}x_{i}=d>0$, so $I_{1}$ and $I_{2}$ cannot
be both empty. Moreover, as long as $d<p$, $I_{2}$ and $I_{3}$
cannot be both empty. Consequently, there are four situations of the
emptiness of the index sets: (1) $I_{1}\ne\varnothing$ and $I_{3}\neq\varnothing$;
(2) $I_{1}\ne\varnothing$, $I_{2}\neq\varnothing$, and $I_{3}=\varnothing$;
(3) $I_{1}=\varnothing$, $I_{2}\neq\varnothing$, and $I_{3}\neq\varnothing$;
and (4) $I_{1}=I_{3}=\varnothing$ and $I_{2}\neq\varnothing$.

Using the same index sets, $y$ can be accordingly divided into three
blocks. By definition (\ref{eq:defn_normal_cone}), it must hold that
\[
\min_{k\in I_{1}}\ y_{k}\ge y_{i+1}=\cdots=y_{j}\ge\max_{k\in I_{3}}\ y_{k}.
\]
Define $I_{U}=\{k\in I_{1}:y_{k}>0\}\cup\{k\in I_{3}:y_{k}<0\}$.
If $I_{U}\neq\varnothing$, then $d_{U}(z)\ge\sqrt{\sum_{k\in I_{U}}y_{k}^{2}}$.
Moreover, $T$ is a hyperplane in $\mathbb{R}^{p}$ with the normal
vector $n_{T}=(1/\sqrt{p},\ldots,1/\sqrt{p})^{\mathrm{T}}$, so $d_{F}(z)=\Vert y\Vert$
and $d_{T}(z)=\Vert y\Vert\cdot|\cos\angle(y,n_{T})|$. We separately
discuss the result based on whether $z\in U$ or $z\notin U$.

\paragraph{The case of $z\in U$}

In this case $d_{U}(z)=0$, so we only need to find $c_{2}$ such
that $c_{2}^{-1}\le\inf_{z\in U}\ |\cos\angle(y,n_{T})|$. Consider
the four situations mentioned above.

(1) $I_{1}\ne\varnothing$ and $I_{3}\neq\varnothing$. Since $x+y=z\in[0,1]^{p}$,
we have $y_{k}\le0$ for $k\in I_{1}$ and $y_{k}\ge0$ for $k\in I_{3}$,
which implies that $y=\mathbf{0}$. Therefore, (\ref{eq:bounded_distance})
holds trivially for any $c_{2}$ since $d_{F}(z)=d_{T}(z)=0$.

(2) $I_{1}\ne\varnothing$, $I_{2}\neq\varnothing$, and $I_{3}=\varnothing$.
We have $\min_{k\in I_{1}}\ y_{k}\ge y_{i+1}=\cdots=y_{p}=c$. Since
$x+y=z\in[0,1]^{p}$, it is true that $0\ge y_{k}\ge c$ for $k\in I_{1}$.
We can assume that $c\neq0$, since otherwise $y=\mathbf{0}$ and
it reduces to the trivial case. Note that $|\cos\angle(y,n_{T})|=|\cos\angle(ty,n_{T})|$
for any $t\neq0$, so we can take $t=1/c$ to obtain
\[
|\cos\angle(ty,n_{T})|=\frac{|\sum_{k=1}^{p}ty_{k}|}{\sqrt{\sum_{k=1}^{p}(ty_{k})^{2}}\cdot\sqrt{p}}=\frac{p-i+\sum_{k\in I_{1}}ty_{k}}{\sqrt{p-i+\sum_{k\in I_{1}}(ty_{k})^{2}}\cdot\sqrt{p}}\ge\frac{\sqrt{p-i+\sum_{k\in I_{1}}ty_{k}}}{\sqrt{p}}.
\]
The inequality holds because $0\le(ty_{k})^{2}\le ty_{k}\le1$ for
$k\in I_{1}$. Using the fact that $i<d$, we immediately get $|\cos\angle(y,n_{T})|=|\cos\angle(ty,n_{T})|\ge\sqrt{(p-i)/p}>\sqrt{1-d/p}$.

(3) $I_{1}=\varnothing$, $I_{2}\neq\varnothing$, and $I_{3}\neq\varnothing$.
In this case $y_{1}=\cdots=y_{j}=c$, and $c\ge y_{k}\ge0$ for $k\in I_{3}$.
Since $x$ needs to satisfy the condition $\sum_{k=1}^{p}x_{k}=\sum_{k=1}^{j}x_{k}=d$
with $0<x_{k}<1$ for $k\le j$, we have $j\ge d+1$. Using the similar
argument in the second case, we take $t=1/c$, and then
\[
|\cos\angle(y,n_{T})|=|\cos\angle(ty,n_{T})|=\frac{j+\sum_{k\in I_{3}}ty_{k}}{\sqrt{j+\sum_{k\in I_{3}}(ty_{k})^{2}}\cdot\sqrt{p}}\ge\sqrt{j/p}\ge\sqrt{(d+1)/p}.
\]

(4) $I_{1}=I_{3}=\varnothing$ and $I_{2}\neq\varnothing$ imply $y_{1}=\cdots=y_{p}=c$
and $|\cos\angle(y,n_{T})|=1$. To summarize, for $z\in U$, we can
choose any $c_{2}$ such that $c_{2}\ge\sqrt{p/(d+1)}$, assuming
$d\le(p-1)/2$.

\paragraph{The case of $z\protect\notin U$}

In this case we assert that $y\neq\mathbf{0}$, and then without loss
of generality we assume that $y_{k}$'s are in decreasing order. Similar
to the discussion above, we consider the four situations based on
the emptiness of $I_{1}$, $I_{2}$, and $I_{3}$.

(1) We have $\min_{k\in I_{1}}\,y_{k}\ge\max_{k\in I_{3}}\,y_{k}$.
Let $M=\{k:|y_{k}|\ge|y_{k'}|,k'\neq k\}$, and then we find that
$M\cap I_{U}\neq\varnothing$. Let $s$ be any element in $M\cap I_{U}$,
and we have $|y_{s}|/\Vert y\Vert\ge1/\sqrt{p}$, indicating that
$d_{U}(z)\ge\sqrt{\sum_{k\in I_{U}}y_{k}^{2}}\ge|y_{s}|\ge d_{F}(z)/\sqrt{p}$.

(2) $y_{1}\ge\cdots\ge y_{i}\ge y_{i+1}=\cdots=y_{p}=c$. (a) If $c\ge0$,
then $y_{1}>0$ and $y_{1}$ has the largest absolute value. (b) If
$c<0$ but $|y_{1}|>|c|$, then $y_{1}$ must be positive and again
it has the largest absolute value. In both cases, we get $d_{U}(z)\ge d_{F}(z)/\sqrt{p}$
based on the argument in (1). (c) If $c<0$, $|y_{1}|\le|c|$, and
$y_{1}\le0$, then same as point (2) in the case of $z\in U$, we
have $|\cos\angle(y,n_{T})|\ge\sqrt{1-d/p}$. (d) At last, let $s$
be an index such that $-c\ge y_{1}\ge\cdots\ge y_{s}\ge0\ge y_{s+1}\ge\cdots\ge y_{i}\ge y_{i+1}=\cdots=y_{p}=c$,
and denote $S_{1}=\sum_{k=1}^{s}y_{k}$, $S_{2}=\sum_{k=1}^{s}y_{k}^{2}$,
and $S_{3}=-\sum_{k=s+1}^{p}y_{k}$. Clearly $\sqrt{S_{2}}\ge S_{1}/\sqrt{s}$.
Since $s\le i\le d\le(p-1)/2$, we have $S_{1}<S_{3}$. Recall that
$d_{U}(z)\ge\sqrt{S_{2}}$, $d_{F}(z)=\Vert y\Vert$, and $d_{T}(z)=|S_{3}-S_{1}|/\sqrt{p}$,
so if $p\ge4$ then
\begin{align*}
 & \sqrt{p}\cdot d_{U}(z)+\sqrt{\frac{p}{d+1}}\cdot d_{T}(z)\ge\sqrt{pS_{2}}+(S_{3}-S_{1})/\sqrt{d+1}\\
\ge & \left\{ \left(\sqrt{p(d+1)/s}-1\right)S_{1}+S_{3}\right\} /\sqrt{d+1}\ge(S_{1}+S_{3})/\sqrt{d+1}\\
\ge & \frac{\Vert y\Vert}{\sqrt{d+1}}\cdot\frac{p-i+\sum_{k\in I_{1}}|y_{k}/c|}{\sqrt{p-i+\sum_{k\in I_{1}}(y_{k}/c)^{2}}}\ge\sqrt{\frac{p-d}{d+1}}\cdot\Vert y\Vert\ge d_{F}(z).
\end{align*}

(3) $y_{1}=\cdots=y_{j}=c\ge y_{j+1}\ge\cdots\ge y_{p}$. In the following
two cases, (a) $c\le0$, and (b) $c>0$ but $|y_{p}|>c$, we would
get $d_{U}(z)\ge d_{F}(z)/\sqrt{p}$ using the argument in (1). For
(c) $c>0$, $|y_{p}|\le c$, and $y_{p}\ge0$, point (3) of the case
$z\in U$ shows that $|\cos\angle(y,n_{T})|\ge\sqrt{(d+1)/p}$. The
remaining possibility is (d) $c>0$, $|y_{p}|\le c$, and $y_{p}<0$.
Let $s$ be an index such that $y_{1}=\cdots=y_{j}=c\ge y_{j+1}\ge\cdots\ge y_{s}\ge0\ge y_{s+1}\ge\cdots\ge y_{p}\ge-c$,
and denote $S_{1}=-\sum_{k=s+1}^{p}y_{k}$, $S_{2}=\sum_{k=s+1}^{p}y_{k}^{2}$,
$S_{3}=\sum_{k=1}^{s}y_{k}$, $S_{4}=\sum_{k=1}^{s}y_{k}^{2}$, and
\[
S=\sqrt{pS_{2}}+|S_{3}-S_{1}|/\sqrt{d+1}\le\sqrt{p}\cdot d_{U}(z)+\sqrt{\frac{p}{d+1}}\cdot d_{T}(z).
\]
We can assume that $S_{2}\le\Vert y\Vert^{2}/p$, since otherwise
we directly get $d_{U}(z)\ge\sqrt{S_{2}}>\Vert y\Vert/\sqrt{p}=d_{F}(z)/\sqrt{p}$.
Using the fact that $s\ge j\ge d+1$, we have $S_{1}\le\sqrt{(p-s)S_{2}}\le\sqrt{1-(d+1)/p}\cdot\Vert y\Vert$.
On the other hand, $S_{3}\ge\sqrt{S_{4}}=\sqrt{\Vert y\Vert^{2}-S_{2}}\ge\sqrt{1-1/p}\cdot\Vert y\Vert>S_{1}$,
so
\[
\sqrt{d+1}\cdot S=\sqrt{p(d+1)S_{2}}-S_{1}+S_{3}\ge\left(\sqrt{\frac{p(d+1)}{p-s}}-1\right)S_{1}+S_{3}\ge S_{1}+S_{3}
\]
as long as $d\ge3$. Note that
\[
\frac{S_{1}+S_{3}}{\Vert y\Vert}=\frac{\sum_{k=1}^{p}|y_{k}|}{\sqrt{\sum_{k=1}^{p}y_{k}^{2}}}=\frac{j+\sum_{k=j+1}^{p}|ty_{k}|}{\sqrt{j+\sum_{k=j+1}^{p}(ty_{k})^{2}}}\ge\sqrt{j+\sum_{k=j+1}^{p}|ty_{k}|}\ge\sqrt{d+1}
\]
for $t=1/c$, and we finally get $S\ge\Vert y\Vert=d_{F}(z)$.

(4) The last case $I_{1}=I_{3}=\varnothing$ is trivial, which completes
the proof.

\subsection{Proof of Corollary \ref{cor:fps_constants}}

First, since
\begin{align*}
|f(X)-f(Y)| & =\left|\mathrm{tr}(S(Y-X))+\lambda(\Vert X\Vert_{1,1}-\Vert Y\Vert_{1,1})\right|\\
 & \le\left|\mathrm{tr}(S(Y-X))\right|+\lambda\left|\Vert X\Vert_{1,1}-\Vert Y\Vert_{1,1}\right|\\
 & \le\Vert S\Vert_{F}\cdot\Vert X-Y\Vert_{F}+\lambda\Vert X-Y\Vert_{1,1}\\
 & \le\Vert S\Vert_{F}\cdot\Vert X-Y\Vert_{F}+\lambda p\Vert X-Y\Vert_{F},
\end{align*}
we find that $f(X)$ is Lipschitz continuous with $L=\Vert S\Vert_{F}+\lambda p$.

Second, part (a) of the assumption is trivial. For part (b), recall
that $g_{2}(X)=-\theta_{p}(X)$. Appendix F of \citet{yang2017richer}
shows that for any $Y\in\bar{G}_{2}$, $\Vert\nabla g_{2}(Y)\Vert_{F}\ge1/\sqrt{s_{0}}$,
where $\nabla g_{2}(Y)$ is any subgradient of $g_{2}$ at $Y$, and
$s_{0}$ is the number of zero eigenvalues of $Y$. Obviously $s_{0}\le p$,
so we get $\rho_{2}=1/\sqrt{p}$. Note that $g_{1}(X)=\theta_{1}(X)-1=g_{2}(I-X)$.
Using the same argument, $\rho_{1}\ge1/\sqrt{s_{1}}$, where $s_{1}$
is the number of eigenvalues equal to one for a matrix $Y\in\bar{G}_{1}\cap\mathcal{X}$.
Since $Y\in\mathcal{X}\Rightarrow\Vert Y\Vert_{F}\le\sqrt{d}$, we
have $s_{1}\le d$, so we can take $\rho_{1}=1/\sqrt{d}$.

Third, it is not hard to show that $d_{\mathcal{F}_{2,3}}(X)=\sqrt{[d_{G_{1}}(X)]^{2}+[d_{G_{2}}(X)]^{2}}\le d_{G_{1}}(X)+d_{G_{2}}(X)$.
Then Theorem \ref{thm:dist_ineq} gives the desired result.

\subsection{Proof of Theorem \ref{thm:batch_fps_convergence}}

The proof mainly follows from \citet{ryu2017proximal}, and our new
result is to give explicit constants instead of the mere rate of convergence
in \citet{ryu2017proximal}. For completeness we include the main
steps of the proof here. We use the notation $\mathbf{X}=(X^{(1)},X^{(2)})$
to denote a collection of two $p\times p$ matrices, and then define
the functions $r(\mathbf{X})=I_{\mathcal{E}}(\mathbf{X})$ and $g(\mathbf{X})=f_{1}(X^{(1)})+f_{2}(X^{(2)})$,
where $I_{\mathcal{E}}(\mathbf{X})=0$ if $X^{(1)}=X^{(2)}\in\mathcal{X}$,
and $I_{\mathcal{E}}(\mathbf{X})=\infty$ otherwise. Since $f_{1}$
and $f_{2}$ are Lipschitz continuous with constants $L_{1}=\lambda p$
and $L_{2}=\Vert S\Vert_{F}+\mu(1+\sqrt{(p+d)(d+1)})$, respectively,
it is easy to show that
\begin{align*}
|g(\mathbf{X})-g(\mathbf{Y})| & \le|f_{1}(X^{(1)})-f_{1}(Y^{(1)})|+|f_{2}(X^{(2)})-f_{2}(Y^{(2)})|\\
 & \le L_{1}\Vert X^{(1)}-Y^{(1)}\Vert_{F}+L_{2}\Vert X^{(2)}-Y^{(2)}\Vert_{F}\\
 & \le\sqrt{L_{1}^{2}+L_{2}^{2}}\cdot\sqrt{\Vert X^{(1)}-Y^{(1)}\Vert_{F}^{2}+\Vert X^{(2)}-Y^{(2)}\Vert_{F}^{2}}\\
 & =\sqrt{L_{1}^{2}+L_{2}^{2}}\cdot\Vert\mathbf{X}-\mathbf{Y}\Vert_{F}.
\end{align*}
Therefore, $g(\cdot)$ is Lipschitz continuous with the constant $L_{g}=\sqrt{L_{1}^{2}+L_{2}^{2}}$.

Denote $\mathbf{X}_{k}=(X_{k},X_{k})$, $\mathbf{Z}_{k}=(Z_{k}^{(1)},Z_{k}^{(2)})$,
and then Algorithm 1 can be equivalently expressed as
\begin{align}
\mathbf{X}_{k+1} & =\mathbf{prox}_{\alpha r}(\mathbf{Z}_{k}),\label{eq:prox_r}\\
\mathbf{Y}_{k+1} & =\mathbf{prox}_{\alpha g}(2\mathbf{X}_{k+1}-\mathbf{Z}_{k}),\label{eq:prox_g}\\
\mathbf{Z}_{k+1} & =\mathbf{Z}_{k}-\mathbf{X}_{k+1}+\mathbf{Y}_{k+1}.\nonumber 
\end{align}
Define the function $p(\mathbf{Z})=(1/\alpha)(\mathbf{X}-\mathbf{Y})$,
where $\mathbf{X}=\mathbf{prox}_{\alpha r}(\mathbf{Z})$ and $\mathbf{Y}=\mathbf{prox}_{\alpha g}(2\mathbf{X}-\mathbf{Z})$,
so we have $p(\mathbf{Z}_{k})=(1/\alpha)(\mathbf{X}_{k+1}-\mathbf{Y}_{k+1})$
and $\mathbf{Z}_{k+1}=\mathbf{Z}_{k}-\alpha p(\mathbf{Z}_{k})$. Let
$X_{*}\in\arg\min_{X\in\mathcal{X}}\,\mathcal{L}(X)$ and denote $\mathbf{X}_{*}=(X_{*},X_{*})$.
Then we have $\mathbf{X}_{*}\in\arg\min_{\mathbf{X}}\,r(\mathbf{X})+g(\mathbf{X})$,
whose optimality condition indicates that $\nabla r(\mathbf{X}_{*})+\nabla g(\mathbf{X}_{*})=\mathbf{O}$,
where $\nabla r(\cdot)$ and $\nabla g(\cdot)$ are some specific
subgradients of $r(\cdot)$ and $g(\cdot)$, respectively. Clearly
we have $\Vert\nabla g(\mathbf{X}_{*})\Vert_{F}=\Vert\nabla r(\mathbf{X}_{*})\Vert_{F}\le L_{g}$.
Moreover, Lemma 1 of \citet{ryu2017proximal} shows that there exists
$\mathbf{Z}_{*}=(Z_{*}^{(1)},Z_{*}^{(2)})$ such that $p(\mathbf{Z}_{*})=\mathbf{O}$
and $\mathbf{X}_{*}=\mathbf{prox}_{\alpha r}(\mathbf{Z}_{*})$.

Next, Lemma 4 of \citet{ryu2017proximal} proves that $\alpha\Vert p(\mathbf{Z})-p(\mathbf{Z}')\Vert_{F}^{2}\le\langle p(\mathbf{Z})-p(\mathbf{Z}'),\mathbf{Z}-\mathbf{Z}'\rangle$
for any $\mathbf{Z}$ and $\mathbf{Z}'$, where $\langle\mathbf{X},\mathbf{Y}\rangle=\mathrm{vec}(\mathbf{X})^{\mathrm{T}}\mathrm{vec}(\mathbf{Y})$.
Therefore,
\begin{align*}
\Vert p(\mathbf{Z}_{k+1})\Vert_{F}^{2} & =\Vert p(\mathbf{Z}_{k})\Vert_{F}^{2}+2\langle p(\mathbf{Z}_{k+1})-p(\mathbf{Z}_{k}),p(\mathbf{Z}_{k})\rangle+\Vert p(\mathbf{Z}_{k+1})-p(\mathbf{Z}_{k})\Vert_{F}^{2}\\
 & =\Vert p(\mathbf{Z}_{k})\Vert_{F}^{2}-2\alpha^{-1}\langle p(\mathbf{Z}_{k+1})-p(\mathbf{Z}_{k}),\mathbf{Z}_{k+1}-\mathbf{Z}_{k}\rangle+\Vert p(\mathbf{Z}_{k+1})-p(\mathbf{Z}_{k})\Vert_{F}^{2}\\
 & \le\Vert p(\mathbf{Z}_{k})\Vert_{F}^{2}-\Vert p(\mathbf{Z}_{k+1})-p(\mathbf{Z}_{k})\Vert_{F}^{2},
\end{align*}
for any $k\ge0$, implying that $\Vert p(\mathbf{Z}_{k})\Vert_{F}^{2}$
is monotonically decreasing. Using the inequality again, we have $\alpha\Vert p(\mathbf{Z}_{k})-p(\mathbf{Z}_{*})\Vert_{F}^{2}=\alpha\Vert p(\mathbf{Z}_{k})\Vert_{F}^{2}\le\langle p(\mathbf{Z}_{k})-p(\mathbf{Z}_{*}),\mathbf{Z}_{k}-\mathbf{Z}_{*}\rangle=\langle p(\mathbf{Z}_{k}),\mathbf{Z}_{k}-\mathbf{Z}_{*}\rangle$,
so
\begin{align*}
\Vert\mathbf{Z}_{k+1}-\mathbf{Z}_{*}\Vert_{F}^{2} & =\Vert\mathbf{Z}_{k}-\mathbf{Z}_{*}\Vert_{F}^{2}-2\alpha\langle p(\mathbf{Z}_{k}),\mathbf{Z}_{k}-\mathbf{Z}_{*}\rangle+\alpha^{2}\Vert p(\mathbf{Z}_{k})\Vert_{F}^{2}\\
 & \le\Vert\mathbf{Z}_{k}-\mathbf{Z}_{*}\Vert_{F}^{2}-\alpha^{2}\Vert p(\mathbf{Z}_{k})\Vert_{F}^{2},
\end{align*}
showing that $\Vert\mathbf{Z}_{k}-\mathbf{Z}_{*}\Vert_{F}^{2}$ is
also monotone. Define $C_{0}=\Vert\mathbf{Z}_{0}-\mathbf{Z}_{*}\Vert_{F}$,
and then $\Vert\mathbf{Z}_{k}-\mathbf{Z}_{*}\Vert_{F}^{2}\le C_{0}^{2}$
and $\Vert\mathbf{Z}_{k}-\mathbf{Z}_{s}\Vert_{F}\le2C_{0}$ for all
$k,s\ge0$. Consequently,
\begin{align}
\sum_{k=0}^{\infty}\Vert p(\mathbf{Z}_{k})\Vert_{F}^{2} & \le\frac{1}{\alpha^{2}}\Vert\mathbf{Z}_{0}-\mathbf{Z}^{*}\Vert_{F}^{2}=\frac{C_{0}^{2}}{\alpha^{2}},\label{eq:sum_pz}\\
\Vert p(\mathbf{Z}_{k})\Vert_{F}^{2} & \le\frac{1}{k}\sum_{k=0}^{\infty}\Vert p(\mathbf{Z}_{k})\Vert_{F}^{2}\le\frac{C_{0}^{2}}{k\alpha^{2}},\label{eq:pz}
\end{align}
where (\ref{eq:pz}) is due to the monotonicity of $\Vert p(\mathbf{Z}_{k})\Vert_{F}^{2}$.

Define $\bar{\mathbf{X}}_{k}=k^{-1}\sum_{j=1}^{k}\mathbf{X}_{k}$,
$\bar{\mathbf{Y}}_{k}=k^{-1}\sum_{j=1}^{k}\mathbf{Y}_{k}$, and $\bar{E}_{k}=g(\bar{\mathbf{Y}}_{k})-g(\mathbf{X}_{*})$.
Equations (29), (31), and (34) of \citet{ryu2017proximal} show that
\begin{align*}
\frac{1}{2}\bar{E}_{k} & \le\frac{1}{2\alpha k}\Vert\mathbf{Z}_{1}-\mathbf{Z}_{*}\Vert_{F}^{2}+\frac{1}{k\alpha}\Vert\mathbf{Z}_{k+1}-\mathbf{Z}_{1}\Vert_{F}\cdot\Vert\nabla r(\mathbf{X}_{*})\Vert_{F},\\
\frac{1}{2}\bar{E}_{k} & \ge\frac{1}{k}\langle\mathbf{Z}_{k}-\mathbf{Z}_{0},\nabla r(\mathbf{X}_{*})\rangle\ge-\frac{1}{k}\Vert\mathbf{Z}_{k}-\mathbf{Z}_{0}\Vert_{F}\cdot\Vert\nabla r(\mathbf{X}_{*})\Vert_{F},
\end{align*}
and then by bounding the relevant terms we get $|\bar{E}_{k}|\le\max\{(C_{0}^{2}+4C_{0}L_{g})/(\alpha k),2C_{0}L_{g}/k\}$.
Moreover,
\[
|g(\bar{\mathbf{X}}_{k})-g(\bar{\mathbf{Y}}_{k})|\le L_{g}\Vert\bar{\mathbf{X}}_{k}-\bar{\mathbf{Y}}_{k}\Vert_{F}=(L_{g}/k)\Vert\mathbf{Z}_{k+1}-\mathbf{Z}_{k}\Vert_{F}\le2C_{0}L_{g}/k,
\]
and then $|g(\bar{\mathbf{X}}_{k})-g(\mathbf{X}_{*})|\le|\bar{E}_{k}|+2C_{0}L_{g}/k$,
implying the first result. The second result is a consequence of Theorem
\ref{thm:unconstrained_problem}(b).

\subsection{Proof of Corollary \ref{cor:stat_error}}

Denote $\mathcal{L}_{*}=\min_{X\in\mathcal{X}}\,\mathcal{L}(X)=\mathcal{L}(\hat{X}_{*})$
and let $\hat{Y}=\mathcal{P}_{\mathcal{K}}(\hat{X})$. Theorem \ref{thm:batch_fps_convergence}
shows that $\Vert\hat{Y}-\hat{X}\Vert_{F}\le C/T$ and $\mathcal{L}(\hat{X})\le\mathcal{L}_{*}+C/T$.
Also Theorem \ref{thm:unconstrained_problem}(b) indicates that $\mathcal{L}(\hat{Y})=f(\hat{Y})\le\mathcal{L}_{*}+C/T\le f(\Pi)+C/T$.

Let $\Delta=\hat{Y}-\Pi$ and $W=S-\Sigma$, and then Lemma 3.1 of
\citet{vu2013fantope} implies that $(\delta/2)\Vert\Delta\Vert_{F}^{2}\le-\mathrm{tr}(\Sigma\Delta)$.
Therefore, if $\lambda\ge\Vert W\Vert_{\infty,\infty}$, then
\begin{align*}
(\delta/2)\Vert\Delta\Vert_{F}^{2} & \le-\mathrm{tr}(\Sigma\Delta)=-\mathrm{tr}(S\Delta)+\mathrm{tr}(W\Delta)\\
 & =f(\hat{Y})-f(\Pi)-\lambda(\Vert\hat{Y}\Vert_{1,1}-\Vert\Pi\Vert_{1,1})+\mathrm{tr}(W\Delta)\\
 & \le_{(*)}f(\hat{Y})-f(\Pi)+2\lambda s\Vert\Delta\Vert_{F}\\
 & \le2\lambda s\Vert\Delta\Vert_{F}+C/T,
\end{align*}
where $(*)$ comes from the proof of Theorem 3.1 of \citet{vu2013fantope}.
Solving the inequality above, we get
\[
\Vert\Delta\Vert_{F}\le\frac{2\lambda s+\sqrt{(2\lambda s)^{2}+2\delta C/T}}{\delta}\le\frac{4\lambda s}{\delta}+\frac{\sqrt{2C/\delta}}{\sqrt{T}},
\]
and hence $\Vert\hat{X}-\Pi\Vert_{F}\le\Vert\Delta\Vert_{F}+\Vert\hat{X}-\hat{Y}\Vert_{F}\le\Vert\Delta\Vert_{F}+C/T$.
Under the stated assumptions, $\Vert W\Vert_{\infty,\infty}\le\lambda$
holds with probability at least $1-2/p^{2}$, thus proving the conclusion.

\subsection{Proof of Theorem \ref{thm:prox_online_fps_convergence} (Part One)}

Define $f_{0t}(X)=-\mathrm{tr}(S_{t}X)$, $f_{1}(X)=\lambda\Vert X\Vert_{1}$,
$f_{2}(X)=\nu\sqrt{pd}[g_{1}(X)]_{+}+\nu p[g_{2}(X)]_{+}$, and $f_{3}(X)=\nu\sqrt{p/(d+1)}d_{C_{1}}(X)$.
Then by Corollary 1, $\nu d_{\mathcal{K}}(X)\le f_{2}(X)+f_{3}(X)$,
so we get $\ell_{t-1}(X)\le(f_{0t}+f_{1}+f_{2}+f_{3})(X)$. Moreover,
$\ell_{t-1}(X)=f_{0t}(X)+f_{1}(X)$ if $X\in\mathcal{K}$. Below we
first follow Proposition 3 of \citet{bertsekas2011incremental} to
obtain inequalities (\ref{eq:ineq_f1}) to (\ref{eq:ineq_f0t}), and
then adapt the results to the online learning setting.

It is easy to see that $X_{t}^{(1)}=\mathbf{prox}_{\alpha_{t}f_{1}}(X_{t}^{(0)})$
and $X_{t}^{(3)}=\mathbf{prox}_{\alpha_{t}f_{3}}(X_{t}^{(2)})$, so
by Proposition 1(b) of \citet{bertsekas2011incremental}, for any
$Y\in\mathcal{X}$ we have
\begin{align}
f_{1}(X_{t}^{(1)})-f_{1}(Y) & \le\frac{1}{2\alpha_{t}}\left(\Vert X_{t}^{(0)}-Y\Vert_{F}^{2}-\Vert X_{t}^{(1)}-Y\Vert_{F}^{2}\right),\label{eq:ineq_f1}\\
f_{3}(X_{t}^{(3)})-f_{3}(Y) & \le\frac{1}{2\alpha_{t}}\left(\Vert X_{t}^{(2)}-Y\Vert_{F}^{2}-\Vert X_{t}^{(3)}-Y\Vert_{F}^{2}\right).
\end{align}
Next, the convexity of $f_{2}$ implies $f_{2}(X_{t}^{(1)})-f_{2}(Y)\le\mathrm{tr}(G(X_{t}^{(1)}-Y))$,
where $G$ is any subgradient of $f_{2}$ at $X_{t}^{(1)}$. Take
$G=\alpha_{t}\nu\sqrt{pd}\mathbf{1}\{\lambda_{1}>1\}\gamma_{1}\gamma_{1}^{\mathrm{T}}-\alpha_{t}\nu p\mathbf{1}\{\lambda_{p}<0\}\gamma_{p}\gamma_{p}^{\mathrm{T}}$,
and then we get
\begin{align*}
\Vert X_{t}^{(2)}-Y\Vert_{F}^{2} & =\Vert X_{t}^{(1)}-\alpha_{t}G-Y\Vert_{F}^{2}=\Vert X_{t}^{(1)}-Y\Vert_{F}^{2}-2\alpha_{t}\mathrm{tr}(G(X_{t}^{(1)}-Y))+\alpha_{t}^{2}\Vert G\Vert_{F}^{2}\\
 & \le\Vert X_{t}^{(1)}-Y\Vert_{F}^{2}-2\alpha_{t}\{f_{2}(X_{t}^{(1)})-f_{2}(Y)\}+\alpha_{t}^{2}\nu^{2}p(p+d),
\end{align*}
indicating that
\begin{equation}
f_{2}(X_{t}^{(1)})-f_{2}(Y)\le\frac{1}{2\alpha_{t}}\left(\Vert X_{t}^{(1)}-Y\Vert_{F}^{2}-\Vert X_{t}^{(2)}-Y\Vert_{F}^{2}\right)+\frac{\alpha_{t}}{2}\nu^{2}p(p+d).
\end{equation}
For $f_{0t}$, we have
\begin{align*}
 & \Vert X_{t}-Y\Vert_{F}^{2}=\Vert\mathcal{P}_{\mathcal{X}}\left(X_{t}^{(3)}+\alpha_{t}S_{t}\right)-Y\Vert_{F}^{2}\\
\le & \Vert X_{t}^{(3)}+\alpha_{t}S_{t}-Y\Vert_{F}^{2}=\Vert X_{t}^{(3)}-Y\Vert_{F}^{2}-2\alpha_{t}\left\{ f_{0t}(X_{t}^{(3)})-f_{0t}(Y)\right\} +\alpha_{t}^{2}\Vert S_{t}\Vert_{F}^{2}
\end{align*}
where the inequality is due to the nonexpansion property of the projection
operator. As a result,
\begin{equation}
f_{0t}(X_{t}^{(3)})-f_{0t}(Y)\le\frac{1}{2\alpha_{t}}\left(\Vert X_{t}^{(3)}-Y\Vert_{F}^{2}-\Vert X_{t}-Y\Vert_{F}^{2}\right)+\frac{\alpha_{t}}{2}\Vert S_{t}\Vert_{F}^{2}.
\end{equation}

Notice that $f_{0t}$, $f_{1}$, $f_{2}$, and $f_{3}$ are all Lipschitz
continuous functions, so $\Vert X_{t}^{(1)}-X_{t}^{(0)}\Vert\le\alpha_{t}\lambda p$,
$\Vert X_{t}^{(2)}-X_{t}^{(1)}\Vert\le\alpha_{t}\nu\sqrt{p(p+d)}$,
and $\Vert X_{t}^{(3)}-X_{t}^{(2)}\Vert\le\alpha_{t}\nu\sqrt{p/(d+1)}$.
Consequently,
\begin{align}
f_{1}(X_{t}^{(0)})-f_{1}(X_{t}^{(1)}) & \le\alpha_{t}(\lambda p)^{2},\\
f_{2}(X_{t}^{(0)})-f_{2}(X_{t}^{(1)}) & \le\alpha_{t}\lambda p\nu\sqrt{p(p+d)},\\
f_{3}(X_{t}^{(0)})-f_{3}(X_{t}^{(3)}) & \le\alpha_{t}\nu\sqrt{p/(d+1)}(\lambda p+\nu\sqrt{p(p+d)}+\nu\sqrt{p/(d+1)}),\\
f_{0t}(X_{t}^{(0)})-f_{0t}(X_{t}^{(3)}) & \le\alpha_{t}\Vert S_{t}\Vert_{F}(\lambda p+\nu\sqrt{p(p+d)}+\nu\sqrt{p/(d+1)}).\label{eq:ineq_f0t}
\end{align}
Adding up (\ref{eq:ineq_f1}) to (\ref{eq:ineq_f0t}), we obtain
\begin{equation}
\ell_{t-1}(X_{t-1})-\ell_{t-1}(Y)\le\frac{\Vert X_{t-1}-Y\Vert_{F}^{2}-\Vert X_{t}-Y\Vert_{F}^{2}}{2\alpha_{t}}+\frac{\alpha_{t}}{2}(\Vert S_{t}\Vert_{F}^{2}+C_{1}\Vert S_{t}\Vert_{F}+C_{2}),\label{eq:loss_one_iteration}
\end{equation}
where $C_{1}=\lambda p+\nu\sqrt{p(p+d)}+\nu\sqrt{p/(d+1)}$ and $C_{2}=\nu^{2}p(p+d)+2(\lambda p)^{2}+2\lambda p\nu\sqrt{p(p+d)}+2\nu\sqrt{p/(d+1)}C_{1}$.
Summarizing (\ref{eq:loss_one_iteration}) over $t=2,\ldots,T+1$,
we have
\begin{align*}
 & \sum_{t=1}^{T}\ell_{t}(X_{t})-\sum_{t=1}^{T}\ell_{t}(Y)\\
\le & \frac{\Vert X_{1}-Y\Vert_{F}^{2}}{2\alpha_{2}}+\frac{1}{2}\sum_{t=2}^{T}(\alpha_{t+1}^{-1}-\alpha_{t}^{-1})\Vert X_{t}-Y\Vert_{F}^{2}+\frac{1}{2}\sum_{t=2}^{T+1}\alpha_{t}(\Vert S_{t}\Vert_{F}^{2}+C_{1}\Vert S_{t}\Vert_{F})+\frac{C_{2}}{2}\sum_{t=2}^{T+1}\alpha_{t}\\
\le & \frac{2d}{\alpha_{2}}+2d\cdot\sum_{t=2}^{T}(\alpha_{t+1}^{-1}-\alpha_{t}^{-1})+\frac{C_{2}}{2}\sum_{t=2}^{T+1}\alpha_{t}+\frac{1}{2}\sum_{t=2}^{T+1}\alpha_{t}(\Vert S_{t}\Vert_{F}^{2}+C_{1}\Vert S_{t}\Vert_{F}).
\end{align*}
Take $\alpha_{1}=\alpha_{0}$, $\alpha_{t}=\alpha_{0}/\sqrt{t-1}$,
$t\ge2$, and then $\sum_{t=2}^{T+1}\alpha_{t}\le2\alpha_{0}\sqrt{T}$
and
\begin{equation}
\sum_{t=1}^{T}\ell_{t}(X_{t})-\sum_{t=1}^{T}\ell_{t}(Y)\le\frac{2d}{\alpha_{0}}+\frac{2d\sqrt{T}}{\alpha_{0}}-\frac{2d}{\alpha_{0}}+\alpha_{0}C_{2}\sqrt{T}+\frac{\alpha_{0}}{2}\sum_{t=1}^{T}\frac{\Vert S_{t+1}\Vert_{F}^{2}+C_{1}\Vert S_{t+1}\Vert_{F}}{\sqrt{t}}.\label{eq:thm4_part1}
\end{equation}

\subsection{Proof of Theorem \ref{thm:prox_online_fps_convergence} (Part Two)}

Let $f_{0}(X)=-\mathrm{tr}(\Sigma X)$ and $\ell=f_{0}+f_{1}+f_{2}+f_{3}$,
and then it is easy to show that
\begin{equation}
\sum_{t=1}^{T}\{\ell(X_{t})-\ell_{t}(X_{t})\}-\sum_{t=1}^{T}\{\ell(\Pi)-\ell_{t}(\Pi)\}=\sum_{t=1}^{T}\mathrm{tr}((S_{t+1}-\Sigma)(X_{t}-\Pi))\coloneqq\sum_{t=1}^{T}\eta_{t}.\label{eq:true_sigma_diff}
\end{equation}
Combining (\ref{eq:thm4_part1}) and (\ref{eq:true_sigma_diff}) yields
\[
\frac{1}{T}\sum_{t=1}^{T}\ell(X_{t})-\ell(\Pi)\le\frac{2d/\alpha_{0}+\alpha_{0}C_{2}}{\sqrt{T}}+\frac{\alpha_{0}}{2T}\sum_{t=1}^{T}\frac{\Vert S_{t+1}\Vert_{F}^{2}+C_{1}\Vert S_{t+1}\Vert_{F}}{\sqrt{t}}+\frac{1}{T}\sum_{t=1}^{T}\eta_{t},
\]
so our target is to bound $\sum_{t=1}^{T}\eta_{t}$, $\sum_{t=1}^{T}\Vert S_{t+1}\Vert_{F}/\sqrt{t}$,
and $\sum_{t=1}^{T}\Vert S_{t+1}\Vert_{F}^{2}/\sqrt{t}$.

First, note that $\eta_{t}\le\Vert S_{t+1}-\Sigma\Vert_{F}\Vert X_{t}-\Pi\Vert_{F}\le2\sqrt{d}\xi_{t+1}$,
so by Assumption \ref{assu:prox_online_fps_assumption},
\[
E\{\exp(u\eta_{t})\}\le E\{\exp(2\sqrt{d}u\xi_{t+1})\}\le\exp(4du^{2}\sigma_{1}^{2}/2),\quad\forall\,|u|\le1/b_{1}.
\]
Also note that $\eta_{t}$ is a martingale difference sequence, so
\begin{align*}
 & E\left\{ \exp\left(u{\textstyle \sum_{t=1}^{T}\eta_{t}}\right)\right\} =E\left\{ \exp\left(u{\textstyle \sum_{t=1}^{T}\eta_{t}}\right)E\left[\exp(u\eta_{T})|S_{1},\ldots,S_{T}\right]\right\} \\
\le & \exp(4du^{2}\sigma_{1}^{2}/2)\cdot E\left\{ \exp\left(u{\textstyle \sum_{t=1}^{T-1}\eta_{t}}\right)\right\} \le\cdots\le\exp(4d\sigma_{1}^{2}Tu^{2}/2),
\end{align*}
showing that $\sum_{t=1}^{T}\eta_{t}$ is sub-exponential with parameters
$b_{T}=b_{1}$ and $\sigma_{1T}=2\sigma_{1}\sqrt{dT}$. Using the
concentration bound for sub-exponential random variables, we have
for any $D_{1}>0$,
\begin{align}
\log\left[P\left\{ \sum_{t=1}^{T}\eta_{t}>D_{1}\sqrt{T}\right\} \right] & \le\left\{ \begin{array}{ll}
-D_{1}^{2}/(8\sigma_{1}^{2}d), & D_{1}\sqrt{T}\le\sigma_{1T}^{2}/b_{1}\\
-D_{1}\sqrt{T}/(2b_{1}), & D_{1}\sqrt{T}>\sigma_{1T}^{2}/b_{1}
\end{array}\right.\nonumber \\
 & \le-\min\left\{ D_{1}^{2}/(8\sigma_{1}^{2}d),D_{1}/(2b_{1})\right\} ,\label{eq:concentration_term1}
\end{align}
where the conservative bound (\ref{eq:concentration_term1}) is used
mainly for brevity.

On the other hand,
\[
\sum_{t=1}^{T}\frac{\Vert S_{t+1}\Vert_{F}}{\sqrt{t}}\le\sum_{t=1}^{T}\frac{\Vert S_{t+1}-\Sigma\Vert_{F}+\Vert\Sigma\Vert_{F}}{\sqrt{t}}\le\sum_{t=1}^{T}\frac{\xi_{t+1}-\mu_{1}}{\sqrt{t}}+2\sqrt{T}(\mu_{1}+\Vert\Sigma\Vert_{F}).
\]
Since $\xi_{t}$ is an independent and sub-exponential sequence, we
have for all $|\lambda|<1/b_{1}$,
\begin{align*}
 & E\left\{ \exp\left(u\sum_{t=1}^{T}\frac{\xi_{t+1}-\mu_{1}}{\sqrt{t}}\right)\right\} =\prod_{t=1}^{T}E\left[\exp\left\{ \frac{u(\xi_{t+1}-\mu_{1})}{\sqrt{t}}\right\} \right]\\
\le & \prod_{t=1}^{T}\exp\left(\frac{u^{2}\sigma_{1}^{2}}{2t}\right)\le\exp\left\{ u^{2}\sigma_{1}^{2}(\log(T)+1)/2\right\} .
\end{align*}
Therefore, for any $D_{2}>0$,
\begin{align}
\log\left[P\left\{ \sum_{t=1}^{T}\frac{\xi_{t+1}-\mu_{1}}{\sqrt{t}}>D_{2}\ell(T)\right\} \right] & \le\begin{cases}
-D_{2}^{2}/(2\sigma_{1}^{2}), & D_{2}\ell(T)\le\sigma_{2T}^{2}/b_{1}\\
-D_{2}\ell(T)/(2b_{1}), & D_{2}\ell(T)>\sigma_{2T}^{2}/b_{1}
\end{cases}\nonumber \\
 & \le-\min\left\{ D_{2}^{2}/(2\sigma_{1}^{2}),D_{2}/(2b_{1})\right\} ,\label{eq:concentration_term2}
\end{align}
where $\ell(T)=\sqrt{\log(T)+1}$, and $\sigma_{2T}=\sigma_{1}\ell(T)$.
With a similar argument, we can show that $\sum_{t=1}^{T}\Vert S_{t+1}\Vert_{F}^{2}/\sqrt{t}\le2\sqrt{T}\mu_{2}+\sum_{t=1}^{T}(\zeta_{t+1}-\mu_{2})/\sqrt{t}$,
and for any $D_{3}>0$, 
\begin{equation}
\log\left[P\left\{ \sum_{t=1}^{T}\frac{\zeta_{t+1}-\mu_{2}}{\sqrt{t}}>D_{3}\ell(T)\right\} \right]\le-\min\left\{ D_{3}^{2}/(2\sigma_{2}^{2}),D_{3}/(2b_{2})\right\} .\label{eq:concentration_term3}
\end{equation}
Let the right hand sides of (\ref{eq:concentration_term1}) (\ref{eq:concentration_term2})
(\ref{eq:concentration_term3}) be $\varepsilon/3$, and we solve
$D_{1}=\max\left\{ 2b_{1}\varepsilon_{l},2\sigma_{1}\sqrt{2d\varepsilon_{l}}\right\} $,
$D_{2}=\max\left\{ 2b_{1}\varepsilon_{l},\sigma_{1}\sqrt{2\varepsilon_{l}}\right\} $,
and $D_{3}=\max\left\{ 2b_{2}\varepsilon_{l},\sigma_{2}\sqrt{2\varepsilon_{l}}\right\} $,
where $\varepsilon_{l}=\log(3/\varepsilon)$. Therefore, with probability
at least $1-\varepsilon$,
\begin{align*}
\frac{1}{T}\mathcal{R}(\{X_{t}\},T) & \le\frac{2d/\alpha_{0}+\alpha_{0}C_{2}+D_{1}}{\sqrt{T}}+\frac{\alpha_{0}C_{1}\left\{ 2\sqrt{T}(\mu_{1}+\Vert\Sigma\Vert_{F})+D_{2}\ell(T)\right\} }{2T}\\
 & \quad+\frac{\alpha_{0}\left\{ 2\sqrt{T}\mu_{2}+D_{3}\ell(T)\right\} }{2T}\\
 & =\frac{C_{3}}{\sqrt{T}}+\frac{C_{4}\{\log(T)+1\}}{T}\coloneqq C(T),
\end{align*}
where $C_{3}=2d/\alpha_{0}+D_{1}+\alpha_{0}\{C_{2}+C_{1}(\mu_{1}+\Vert\Sigma\Vert_{F})+\mu_{2}\},$and
$C_{4}=\alpha_{0}(C_{1}D_{2}+D_{3})/2$. By the convexity of $\ell(\cdot)$,
we have $T^{-1}\sum_{t=1}^{T}\ell(X_{t})\ge\ell(\hat{X}_{T})$, so
with the specified probability, $\ell(\hat{X}_{T})-\ell(\Pi)\le C(T)$.

Let $\hat{Y}_{T}=\mathcal{P}_{\mathcal{K}}(\hat{X}_{T})$. If $\nu\ge\lambda p+\Vert\Sigma\Vert_{F}+1$,
then by Theorem 1, we have $\ell(\hat{Y}_{T})-\ell(\Pi)\le C(T)$
and $\Vert\hat{X}_{T}-\hat{Y}_{T}\Vert\le C(T)$. Let $\Delta=\hat{Y}_{T}-\Pi$,
and then Lemma 3.1 of \citet{vu2013fantope} shows that $(\delta/2)\Vert\Delta\Vert_{F}^{2}\le-\mathrm{tr}(\Sigma\Delta)$,
thus $(\delta/2)\Vert\Delta\Vert_{F}^{2}\le-\mathrm{tr}(\Sigma\Delta)=\ell(\hat{Y}_{T})-\ell(\Pi)-\lambda(\Vert\hat{X}_{T}\Vert_{1,1}-\Vert\Pi\Vert_{1,1})\le C(T)+\lambda\Vert\Pi\Vert_{1,1}$.
Since $\Pi$ is sparse, we have $\Vert\Pi\Vert_{1,1}\le s\Vert\Pi\Vert_{F}=s\sqrt{d}$.
Finally, applying the triangle inequality yields the requested result.

\subsection{Proof of Theorem \ref{thm:mirror_online_fps_convergence}}

To simplify the notation, in this proof we use $\Vert\cdot\Vert_{q}$
as a shorthand for the $\Vert\cdot\Vert_{q,q}$ norm, and let $\mathbb{S}^{p}$
be the space of $p\times p$ symmetric matrices. We first show that
the function $\rho(X)=\Vert X\Vert_{r}^{2}/2$ is $\beta$-strongly
convex with respect to the $\Vert\cdot\Vert_{1}$ norm. To see this,
by Lemma 9 of \citet{kakade2012regularization}, we have $\rho(X)\ge\rho(Y)+\mathrm{tr}\left(U(X-Y)\right)+\{(r-1)/2\}\Vert X-Y\Vert_{r}^{2}$
for all $X,Y\in\mathbb{S}^{p}$ and $U\in\partial\rho(Y)$. In general,
for $0<m<n$ we have $\Vert X\Vert_{m}\le p^{2/m-2/n}\Vert X\Vert_{n}$,
so $\Vert X\Vert_{1}\le p^{2-2/r}\Vert X\Vert_{r}=\exp(2)\Vert X\Vert_{r}$.
Then we immediately get $\rho(X)\ge\rho(Y)+\mathrm{tr}\left(U(X-Y)\right)+(\beta/2)\Vert X-Y\Vert_{1}^{2}$.

Next we verify that $\mathring{f}(X;Y,t)=-\mathrm{tr}(YX)+\lambda t\Vert X\Vert_{1}+\sqrt{t}\Vert X\Vert_{r}^{2}/2$
is Lipschitz continuous on $\mathcal{X}$ with the Lipschitz constant
$L_{t}$. It is easy to show that the first two terms have Lipschitz
constants $\Vert Y\Vert_{F}$ and $\lambda tp$, respectively. For
the third term,
\begin{equation}
\frac{\partial(\Vert X\Vert_{r}^{2}/2)}{\partial x_{kl}}=\frac{1}{r}\left(\sum_{i,j}|x_{ij}|^{r}\right)^{2/r-1}r|x_{kl}|^{r-1}\mathrm{sign}(x_{kl}),\label{eq:r_norm_derivative}
\end{equation}
implying that $\Vert\nabla\rho(X)\Vert_{F}=\Vert X\Vert_{r}^{2-r}\cdot\Vert X\Vert_{2r-2}^{r-1}\le\exp(-4)p^{2}\cdot\Vert X\Vert_{F}$.
So the Lipschitz constant for the third term is $\exp(-4)\sqrt{td}p^{2}$.
Adding the constants together yields the required result.

Using the notation in \citet{orabona2015generalized}, define $F(X)=\lambda\Vert X\Vert_{1}$,
$g(X)=\rho(X)=\Vert X\Vert_{r}^{2}/2$, and $f_{t}(X)=\sqrt{t}g(X)+tF(X)$.
The domain of $f_{t}(X)$ is taken to be $\mathcal{K}$. From the
first result above, it is obvious that $f_{t}(X)$ is $\beta\sqrt{t}$-strongly
convex with respect to $\Vert\cdot\Vert_{1}$. Let $f_{t}^{*}(Y)=\sup_{X\in\mathcal{K}}\,\{\mathrm{tr}(YX)-f_{t}(X)\}$
be the Fenchel conjugate of $f_{t}$, with $\mathbb{S}^{p}$ as the
domain. \citet{orabona2015generalized} shows that $f_{t}^{*}$ is
everywhere differentiable, and $\nabla f_{t}^{*}(Y)=\arg\min_{X\in\mathcal{K}}\,\{-\mathrm{tr}(YX)+f_{t}(X)\}$.

Let $W_{t}=\nabla f_{t}^{*}(Y_{t})=\arg\min_{X\in\mathcal{K}}\,\{-\mathrm{tr}(Y_{t}X)+f_{t}(X)\}$,
and then Theorem 1(a) indicates that $\mathring{\mathcal{L}}_{*}\coloneqq\min_{X\in\mathcal{X}}\,\mathring{\mathcal{L}}(X;Y_{t},t)=\mathring{\mathcal{L}}(W_{t};Y_{t},t)$.
Also let $X_{t}$ be defined as in Algorithm 3, so by definition,
$\mathring{\mathcal{L}}(X_{t};Y_{t},t)\le\mathring{\mathcal{L}}(W_{t};Y_{t},t)+\beta\sqrt{t}\varepsilon_{t}^{2}/2$.
In this sense, $X_{t}$ is an approximation to $W_{t}$. In fact,
by the strong convexity of $\mathring{\mathcal{L}}(X;Y_{t},t)$ with
respect to the $\Vert\cdot\Vert_{1}$ norm, we have $\beta\sqrt{t}\varepsilon_{t}^{2}/2\ge\mathring{\mathcal{L}}(X_{t};Y_{t},t)-\mathring{\mathcal{L}}_{*}\ge(\beta\sqrt{t}/2)\Vert X_{t}-W_{t}\Vert_{1}^{2}$.
Therefore, we assert that $\Vert X_{t}-W_{t}\Vert_{1}\le\varepsilon_{t}$,
and consequently $\mathrm{tr}((W_{t}-X_{t})S_{t+1})\le\psi_{t+1}\Vert X_{t}-W_{t}\Vert_{1}\le\varepsilon_{t}\psi_{t+1}$.

Next, Lemma 1 of \citet{orabona2015generalized} shows that
\begin{equation}
\sum_{t=1}^{T}\mathrm{tr}\left(S_{t+1}(\Pi-W_{t})\right)\le f_{T}(\Pi)+\sum_{t=1}^{T}\left\{ \frac{\psi_{t+1}^{2}}{2\beta\sqrt{t}}+f_{t-1}(W_{t})-f_{t}(W_{t})\right\} .\label{eq:omd_lemma}
\end{equation}
Then adding the $\mathrm{tr}((W_{t}-X_{t})S_{t+1})$ term to the left
hand side of (\ref{eq:omd_lemma}) yields
\begin{align}
\sum_{t=1}^{T}\mathrm{tr}\left(S_{t+1}(\Pi-X_{t})\right) & \le f_{T}(\Pi)+\sum_{t=1}^{T}\left\{ \frac{\psi_{t+1}^{2}}{2\beta\sqrt{t}}+\varepsilon_{t}\psi_{t+1}+(\sqrt{t-1}-\sqrt{t})g(W_{t})-F(W_{t})\right\} \nonumber \\
 & \le\sqrt{T}g(\Pi)+TF(\Pi)+\sum_{t=1}^{T}\left\{ \frac{\psi_{t+1}^{2}}{2\beta\sqrt{t}}+\varepsilon_{t}\psi_{t+1}-F(W_{t})\right\} .\label{eq:omd_regret}
\end{align}
Note that $|F(W_{t})-F(X_{t})|=\lambda|\Vert W_{t}\Vert_{1,1}-\Vert X_{t}\Vert_{1,1}|\le\lambda\Vert W_{t}-X_{t}\Vert_{1,1}\le\lambda\varepsilon_{t}$,
so adding the inequality $0\le F(W_{t})-F(X_{t})+\lambda\varepsilon_{t}$
to (\ref{eq:omd_regret}) gives
\[
\sum_{t=1}^{T}\mathrm{tr}\left(S_{t+1}(\Pi-X_{t})\right)+\sum_{t=1}^{T}F(X_{t})-TF(\Pi)\le\sqrt{T}g(\Pi)+\sum_{t=1}^{T}\left\{ \frac{\psi_{t+1}^{2}}{2\beta\sqrt{t}}+\varepsilon_{t}(\lambda+\psi_{t+1})\right\} .
\]
Finally, $d_{\mathcal{K}}(\Pi)=0$, and $d_{\mathcal{K}}(X_{t})\le\beta\sqrt{t}\varepsilon_{t}^{2}/2$
by Theorem \ref{thm:unconstrained_problem}(b), so the first part
of the theorem is proved.

Now consider the final output $\hat{X}_{T}=X_{T}$. Define $Z_{T}=\mathcal{P}_{\mathcal{K}}(X_{T})$,
$S=T^{-1}\sum_{t=1}^{T}S_{t}$, $\mathcal{L}(X)=T^{-1}\mathring{\mathcal{L}}(X;\alpha TS,T)$,
and $\mathcal{L}_{*}=\min_{X\in\mathcal{X}}\,\mathcal{L}(X)$. Then
by Algorithm 3 and Theorem 1, we have $\mathcal{L}(Z_{T})\le\mathcal{L}_{*}+\beta\varepsilon_{T}^{2}/(2\sqrt{T})$
and $\Vert X_{T}-Z_{T}\Vert_{F}\le\beta\sqrt{T}\varepsilon_{T}^{2}/2$.
Define $\Delta=Z_{T}-\Pi$ and $W=S-\Sigma$. Similar to the proof
of Corollary \ref{cor:stat_error}, if $\lambda\ge\Vert W\Vert_{\infty}$,
then
\begin{align*}
(\delta_{d}/2)\Vert\Delta\Vert_{F}^{2} & \le-\mathrm{tr}(\Sigma\Delta)=-\mathrm{tr}(S\Delta)+\mathrm{tr}(W\Delta)\\
 & =\mathcal{L}(Z_{T})-\mathcal{L}(\Pi)-\lambda(\Vert Z_{T}\Vert_{1}-\Vert\Pi\Vert_{1})+\mathrm{tr}(W\Delta)-(\rho(Z_{T})-\rho(\Pi))/\sqrt{T}\\
 & \le\beta\varepsilon_{T}^{2}/(2\sqrt{T})+2\lambda s\Vert\Delta\Vert_{F}-(\rho(Z_{T})-\rho(\Pi))/\sqrt{T}.
\end{align*}
Due to the strong convexity of $\rho(X)$, we have $\rho(Z_{T})-\rho(\Pi)\ge\mathrm{tr}\left(U\Delta\right)+(\beta/2)\Vert\Delta\Vert_{1}^{2}\ge-\Vert U\Vert_{F}\Vert\Delta\Vert_{F}+(\beta/2)\Vert\Delta\Vert_{F}^{2}$,
where $U=\nabla\rho(\Pi)$. Since $\Pi$ is sparse, the norm of $\Pi$
can be computed on an $s\times s$ submatrix. Therefore, using (\ref{eq:r_norm_derivative})
again we get $\Vert\nabla\rho(\Pi)\Vert_{F}=\Vert\Pi\Vert_{r}^{2-r}\cdot\Vert\Pi\Vert_{2r-2}^{r-1}\le s^{2-4/\log(p)}\Vert\Pi\Vert_{F}=s^{2-4/\log(p)}\sqrt{d}$.
Further take $\varepsilon_{T}=1/\sqrt{T}$, and then
\[
(\delta_{d}+\beta/\sqrt{T})/2\cdot\Vert\Delta\Vert_{F}^{2}\le\frac{\beta}{2T^{3/2}}+\left(2\lambda s+\frac{\Vert U\Vert_{F}}{\sqrt{T}}\right)\Vert\Delta\Vert_{F}.
\]
Solving the inequality and noting that $\Vert X_{T}-Z_{T}\Vert_{F}\le\beta\sqrt{T}\varepsilon_{T}^{2}/2=\beta/(2\sqrt{T})$,
we get the claimed bound.

\singlespacing

\bibliographystyle{apalike}
\bibliography{ref}

\end{document}